\documentclass[aps,reprint,superscriptaddress,floatfix,showpacs,pra]{revtex4-2}
\usepackage{graphicx}
\usepackage{amsmath}
\usepackage{amssymb}
\usepackage{amsfonts}

\usepackage[unicode=true,breaklinks=true,colorlinks=true]{hyperref}
\hypersetup{citecolor=blue,urlcolor=blue}

\usepackage{newtxtext}
\usepackage{newtxmath}
\usepackage[all]{hypcap}
\usepackage{natbib}
\usepackage{multirow}
\usepackage{booktabs}
\usepackage{bbold}
\usepackage{bm}
\usepackage{bbm}

\newcommand{\ket}[1]{|{#1}\rangle}

\begin{document}
	
	\title{Bayesian-based  hybrid method for rapid optimization of NV center sensors}
	\author{Jiazhao Tian}
	\email{tianjiazhao@tyut.edu.cn}
	\affiliation{School of physics, Taiyuan university of technology, Taiyuan 430000 P. R. China}
	\author{Ressa S. Said}
	\affiliation{Institute for Quantum Optics and Center for Integrated Quantum Science and Technology, Ulm University, D-89081 Ulm, Germany}
	\author{Fedor Jelezko}
	\affiliation{Institute for Quantum Optics and Center for Integrated Quantum Science and Technology, Ulm University, D-89081 Ulm, Germany}
	\author{Jianming Cai}
	\affiliation{School of Physics, International Joint Laboratory on Quantum Sensing and Quantum Metrology, Huazhong University of Science and Technology, Wuhan 430074 P. R. China}
	\affiliation{State Key Laboratory of Precision Spectroscopy, East China Normal University, Shanghai, 200062, China}
	\author{Liantuan Xiao}
	\affiliation{School of physics, Taiyuan university of technology, Taiyuan 430000 P. R. China}
	
	\date{\today}

\begin{abstract}
	NV center is one of the most promising platforms in the field of quantum sensing. Magnetometry based on NV center, especially, has achieved a concrete development in regions of biomedicine and medical diagnostics. Improving the sensitivity of NV center sensor under wide inhomogeneous broadening and filed amplitude drift is one crucial issue of continuous concern, which relies on the coherent control of NV center with higher average fidelity. Quantum optimal control (QOC) methods provide access to this target, nevertheless the high time consumption of current methods due to the large number of needful sample points as well as the complexity of the parameter space has hindered their usability. In this paper we propose the Bayesian estimation phase-modulated (B-PM) method to tackle this problem. In the case of state transforming of NV center ensemble, the B-PM method reduces the time consumption by more than $90\%$ compared to the conventional standard Fourier base (SFB) method while increasing the average fidelity from $0.894$ to $0.905$. In AC magnetometry scenery, the optimized control pulse given by B-PM method achieves a eight-fold extension of the coherence time $T_2$  compared to rectangular $\pi$ pulse. Similar application can be made in other sensing situations. As a general algorithm, the B-PM method can be further extended to open- and closed-loop optimization of complex systems based on a variety of quantum platforms.
\end{abstract}

\maketitle
	
	\section{Introduction}
	
	Nitrogen-vacancy (NV) center in diamond shows its bright prospect in quantum sensing of magnetic field\cite{taylorHighsensitivityDiamondMagnetometer2008,rondinMagnetometryNitrogenvacancyDefects2014,casolaProbingCondensedMatter2018}, electric field \cite{doldeElectricfieldSensingUsing2011a}, temperature\cite{neumannHighPrecisionNanoscaleTemperature2013a,PhysRevApplied.10.034009} and strain \cite{dohertyElectronicPropertiesMetrology2014d}. Among these areas, researches on NV center based ultrasensitive magnetometry have achieved fast development\cite{wangPicoteslaMagnetometryMicrowave2022,schmittSubmillihertzMagneticSpectroscopy2017} and are on the road to practical and commercial applications in biomedicine and diagnostics\cite{millerSpinenhancedNanodiamondBiosensing2020,liSARSCoV2QuantumSensor2022,araiMillimetrescaleMagnetocardiographyLiving2022,chenImmunomagneticMicroscopyTumor2022}.
	The inhomogeneous broadening due to ambient nuclear spins and external bias field is one of the main obstacle to further improving the sensitivity of the sensor. To alleviate the problem, dynamic decoupling (DD) is widely applied in sensing strategies based on NV center\cite{wangRandomizationPulsePhases2019, macquarrieContinuousDynamicalDecoupling2015,caoProtectingQuantumSpin2020b}, and varies of optimal method are used to increase the performance of DD sequence\cite{farfurnikOptimizingDynamicalDecoupling2015a,genovEfficientRobustSignal2020a,poulsenOptimalControlNitrogenvacancy2022}, since the inhomogeneous broadening will also damages the fidelity of control pulses with finite pulse length that construct the DD series. Adiabatic strategies can significantly prolong the decoherence time $T_2$ and sensitivity compared to the conventional flat $\pi$ pulses\cite{genovEfficientRobustSignal2020a}, while smooth shaped pulse design based on numerical optimization\cite{poulsenOptimalControlNitrogenvacancy2022} can adapt to a wider pulse length range where adiabatic condition is not satisfied. One major drawback of numerical optimal design is its time efficiency. On one hand, multiple sample points need to be measured to give a accurate value of the objective function denoting the average fidelity over the frequency and field amplitude broadening, such the processing time of calling the objective function once is prolonged. On the other hand, the total calling times of the objective function will increase along with the number of parameters, while in most cases a larger number of parameter (over 10) is needful to guarantee a fair fidelity. A rapid while well-performed optimization method is anticipated to improve the usability of such control strategies.
	
	In this work we propose the Bayesian estimation phase-modulated (B-PM) method to overcome the time consumption problem. Unlike the developed Bayesian methods\cite{brochuTutorialBayesianOptimization2010, shahriariTakingHumanOut2016}, the B-PM method grafts the Bayesian estimation model onto the direct search method, such circumvent the complex process to calculate the acquisition functions\cite{zhanExpectedImprovementExpensive2020}. Further taking advantage to the phase-modulated method, the B-PM method makes itself a efficient hybrid optimization method for robust quantum control against noise.
	The objective function of optimization process under consideration is the average value of multiple function with different detuning and amplitude value of the control filed. The value of these multiple function can specifically refer to the fidelity between the final and target states, gate operators and so on.   Using Bayesian based estimation model, we  make an accurate prediction of the average fidelity based on a small number of sample points, such the computation time of calling the objective function once is reduced. In addition, adopting phase-modulated base allow the control field comprising multiple frequency components with less parameters, which lead to a significant decrease of the necessary total calling number of the objective function to find the local optimal results. We further verify the computation time of estimation process is negligible compared with the sample measuring process. Overall the B-PM method decrease the total time consumed during the entire optimization process. We firstly apply the B-PM method to the state flipping of NV center ensemble. Comparing with the conventional standard Fourier bases (SFB) method without estimation, the B-PM method increase the average fidelity from $0.894$ to $0.905$ with only $9.3\%$ total time consumption. When applied to the sensing strategy of AC magnetic signal, the B-PM shaped pulse prolong the decoherence time $T_2$ by $8$ times compared to the conventional rectangular $\pi$ pulse. The B-PM method can be extended to DC sensing strategy\cite{bauchUltralongDephasingTimes2018}, and the open- and closed-loop optimization process for open systems\cite{glaserTrainingSchrodingerCat2015,kochQuantumOptimalControl2022} and many-body systems\cite{accantoRapidRobustControl2017,yangProbeOptimizationQuantum2020,eggerAdaptiveHybridOptimal2014a}.
	%%%%%%%%%%%%%%%%%%%%%%%%%%%%%%%%%%%%%%%%%%
	\section{Methods}
	
	\subsection{Optimal control model of NV center ensemble}

	\begin{figure*}[]
		\centering 
		\includegraphics[width = 15cm]{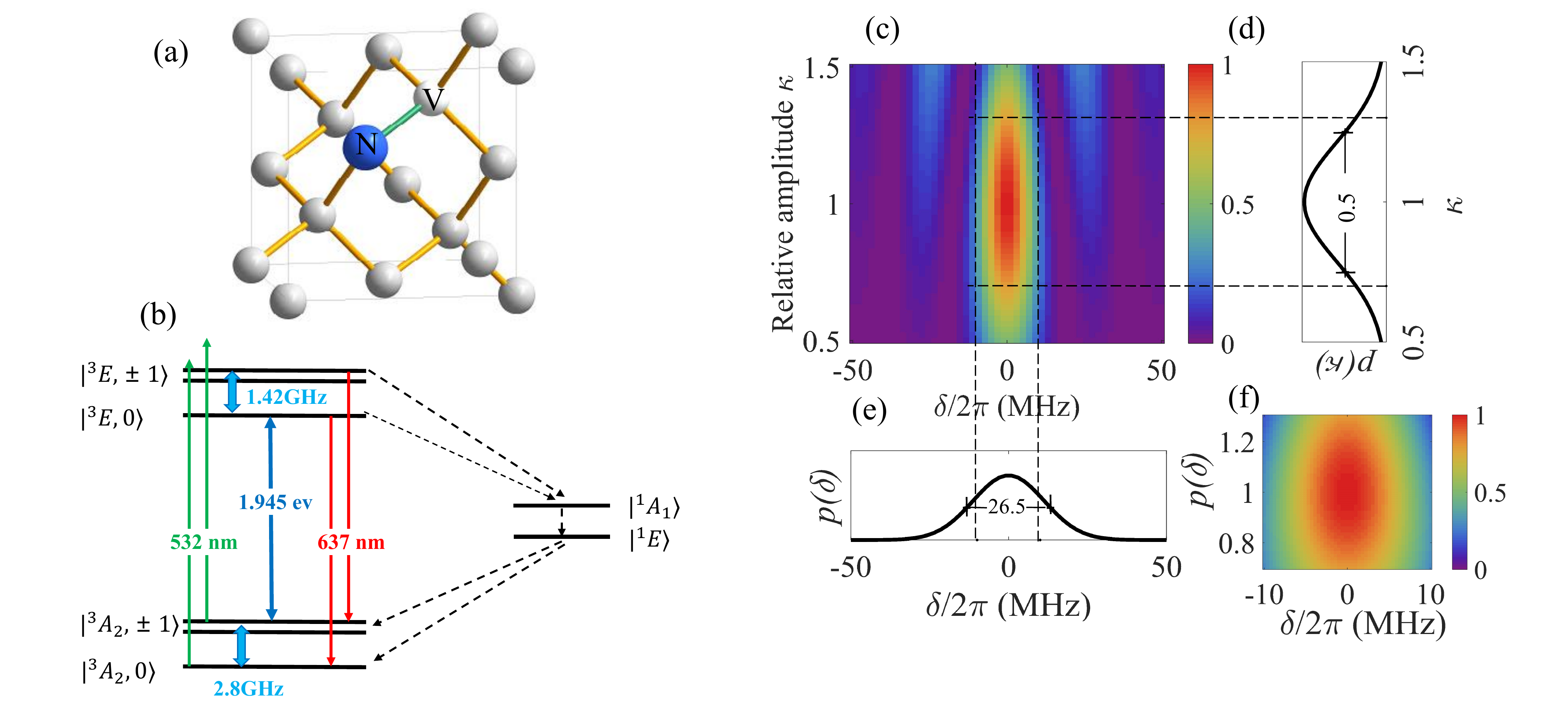}
		\caption{ (\textbf{a}) Schematic of NV center in diamond lattice. (\textbf{b}) Schematic diagram of the energy level of NV center. \textbf{(c)} Fidelity of state flip of NV center using rectangular control field $g(t) = 10$ MHz. (\textbf{d}) The probability density of frequency detuning $\delta$. (\textbf{e}) The probability density of amplitude drift factor $\delta$. (\textbf{f}) The sampling range of numerical simulating when computing the value of average fidelity $\mathcal{F}$.}\label{Fig_1}
	\end{figure*}
	The nitrogen-vacancy center (NV center) in diamond\cite{jelezkoSingleDefectCentres2006} has an triplet ground state with electron spin $S = 1$ and zero-field splitting $D = 2\pi\times2.88$ GHz. The energy gap between ground and excited state is $1.945$ ev (637 nm), and due to uneven radiation probability from excited state $|m_s = \pm1\rangle$ and $|m_s = 0\rangle$ to the radiationless metastable intermediate state, NV center can be optically initialized and read out. The structure of NV center and energy level scheme is shown in Fig. \ref{Fig_1} (a) and (b). The Hamiltonian of the triplet ground state of NV center can be expressed as
	\begin{equation}
		H=D\left(S_Z^2-\frac{2}{3}\right)+\gamma \mathbf{B}\cdot \mathbf{S}+H_{\text{ele}}+H_{\text{hf}},
	\end{equation}
	where $D$ is the zero-field splitting, $\mathbf{B}$ is the magnetic vector, $\gamma_{\text{NV}} = 2\pi\times2.8$ MHz/G is the gyromagnetic ratio of NV center, $H_{\text{ele}}$ is the electric interaction term with coupling coefficient $\sim$ Hz/(V/m) and $H_{\text{hf}}$ is the hyperfine coupling term express interactions between NV center and ambient nuclear spins. Under the magnetic field along $z$ direction $\mathbf{B} = B_z$ and neglecting the electric coupling term, the Hamiltonian of the two-level subspace formed by the $m_s = 0$ ground state $|\psi_0\rangle$ and the $m_s = -1$ (or $m_s = 1$) ground state $|\psi_1\rangle$ can be written as
	\begin{equation}\label{Eq: H single_2level}
		H=\frac{D+\gamma B_z}{2}\sigma_{z}.
	\end{equation}

	One simple but requisite control target is to flip all spins in the NV ensemble from one state to another with high average fidelity. Consider an ensemble of two-level system described by equation \ref{Eq: H single_2level} which is controlled by a time-dependent field $g(t)$, the Hamiltonian of each spin in the ensemble can be represented as
	\begin{equation}
		H(t)=\frac{\omega_{0}+\delta}{2} \sigma_{z}+\kappa g(t) \sigma_{x},
	\end{equation}
	with $\omega_0 = D+\gamma B_z $ the unperturbed energy gap between $|\psi_0\rangle$  and $|\psi_1\rangle$, $\delta$ the normally distributed detuning term as a result of inhomogeneous local ambient among the ensemble, and $\kappa$ the amplitude drift factor varies with different control trials. The probability density of $\delta$ and $\kappa$ can be noted as
	\begin{equation}
		p(\delta)=\frac{1}{\sqrt{2 \pi} \sigma_{\delta}} e^{-\frac{\delta^2}{2 \sigma_{\delta}^2}}
	\end{equation}
	and
	\begin{equation}
		p(\kappa)=\frac{1}{\sqrt{2 \pi} \sigma_{\kappa}} e^{-\frac{\kappa^2}{2 \sigma_{\kappa}^2}}
	\end{equation}
	respectively, where we take the full width at half maximum (FWHM) of $p(\delta)$ as $2\pi \times 26.5$MHz, corresponding to a dephasing time of $T_2^*\approx 20$ns \cite{genovEfficientRobustSignal2020a}, and FWHM of $p(\kappa)$ as $0.5$. Figure~\ref{Fig_1} (c-e) depict the fidelity
	\begin{equation}\label{Eq:single_f}
		f(\delta,\kappa) = \left|\left\langle\psi_g \mid \psi(\delta,\kappa,T)\right\rangle\right|^2
	\end{equation}
	as a function of detuning $\delta$ and amplitude drift $\kappa$ of the rectangular control field $g(t) = \pi/T = 2\pi\times10$ MHz, where $T$ is the evolution time, $\psi(\delta,\kappa,T)$ is the final state, and the initial state and the target state is taken as $\psi(\delta,\kappa,0) = \psi_{0}$ and $\psi_g = \psi_{1}$ respectively. Based on Figure~\ref{Fig_1} (c) we take the sample range as $\delta \in 2\pi \times [-10,10]$ MHz and $\kappa\in [0.5,1.5]$, as is showed in Figure~\ref{Fig_1} (f), and the optimization target is to improve the average fidelity among this range.

	Taking into account the inhomogeneous detuning as well as the amplitude drift, the explicit form of the average fidelity can be represented as
	\begin{equation}
		\mathcal{F}=\iint d \delta d \kappa p(\delta) p(\kappa) f(\delta,\kappa).
	\end{equation}
	For fixed $\delta, \kappa$ and $T$, $\psi(\delta,\kappa,T)$ is functional of control field $g(t)$, therefore is the function of parameters $\bm{\lambda}$ that construct $g(t)$. For example, $g(t)$ can be constructed based on the standard Fourier basis as 
	\begin{equation}\label{Eq:g_SFB}
		g_{\mathrm{SFB}}(t)=\sum_{j=1}^{N_{\text{D}}} a_j \cos \left(\omega_j t+\phi_j\right) \cos \left(\omega_0 t\right)
	\end{equation}
	with $\bm{\lambda} = \{\bm{a},\bm{b},\bm{\phi}\}$, or on the phase modulated Fourier basis\cite{tianQuantumOptimalControl2020} as 
	\begin{equation}\label{Eq:g_PM}
		g_{\mathrm{PM}}(t)=\sum_{j=1}^{N_{\text{D}}} a_j\cos \left[\omega_0 t+\frac{b_j}{\nu_j} \sin \left(\nu t\right)\right]
	\end{equation}
	with $\bm{\lambda} = \{\bm{a},\bm{b},\bm{\nu}\}$. The optimization target, therefore, is to find the parameters $\bm{\lambda}$ that gives the highest value of $\mathcal{F}$ while the maximum amplitude of $g(t)$ is limited to $g_{\text{max}}$. We concisely describe this optimization model as
	\begin{equation}\label{Eq:optim_model}
		\begin{array}{ll}
			\min & 1-\mathcal{F} \\
			\text { s.t. } & g(\boldsymbol{\bm{\lambda}})-g_{\text{max}} \leq0.
		\end{array}
	\end{equation}

	Plenty of optimization methods can be used to solve this problem, with $\bm{\lambda}$ being the optimization parameters and $\mathcal{F}$ being the objective function. Ideally, the value of $\mathcal{F}$ should be calculated by averaging the fidelity of a large number of samples taken from the whole ensemble with several repeated control processes. In practice, this time-consuming process is usually replace by taken the weighted average of several equidistant sampling among certain range and can be represented as
	\begin{equation}\label{Eq:Fobj}
		\mathcal{F}_{\text {obj }}=\mathcal{N} \sum_{k=1}^M \sum_{j=1}^N p\left(\delta_k\right) p\left(\kappa_j\right) f\left(\delta_k,\kappa_j\right)
	\end{equation}
	wherein the distribution of $\delta$ and $\kappa$ is assumed to be known, $M, N$ are the number of different values of $\delta$ and $\kappa$ respectively,  and $\mathcal{N}=\left[\sum_{k=1}^M \sum_{j=1}^N p\left(\delta_k\right) p\left(\kappa_j\right)\right]^{-1}$ is the normalization constant. Even so, this process still consume a substantial amount of time. Below we will show how the Bayesian-based estimation method can be used in the optimization process and combined with the PM method to efficiently reduce the total execution time.

	\subsection{The estimation model}
	One essential task of the optimal process is to build an computationally cheap estimation model, also known as the metamodel\cite{simpsonMetamodelsComputerbasedEngineering2001,wangReviewMetamodelingTechniques2006,r.r.bartonMetamodelingStateArt11}, of the expensive function. Bayesian statics and analysis methods provide a powerful tool to complete this task. From the viewpoint of Bayesian statistics, any event can be endowed with a probabilistic property, whether or not it is a stochastic event. So the undetermined parameter $\theta$ in a static model can be taken as a realization of stochastic process $\bm{\theta}$ subject to a prior distribution $p(\theta)$, which represents the available knowledge or simply beliefs about $\bm{\theta}$ before any observation of samples. This prior knowledge need to be updated with the information in observed sample data $\bm{s}$, to give the final description about $\bm{\theta}$ represent by the posterior distribution $p(\theta|\bm{s})$. This relationship is explicitly formulated by the Bayesian theorem,
	\begin{equation}
		p(\theta|\bm{s}) = \frac{\mathcal{L}(\theta|\bm{s})p(\theta)}{p(\bm{s})},
	\end{equation}
	wherein the likelihood function $\mathcal{L}(\theta|\bm{s}) = p(\bm{s}|\theta)$ is defined as the conditional probability distribution of the given parameters of the data \cite{vandeschootBayesianStatisticsModelling2021}.

	We limit the considered target true functions to be deterministic, which means all trials with the same input parameters gives identical function value. Follow the principle of Bayesian statistic, the deterministic response function $y(\bm{x})$ of a $k$-dimension variable $\bm{x} = \{x_1, x_2, \cdots, x_k\}'$ can be treated as a realization of stochastic process $Y(\bm{x})$ \cite{jeromesacksDesignAnalysisComputer1989}:
	\begin{equation}\label{Eq:Y_x}
		Y(\bm{x}) =\sum_{h}\beta_h f_h(\bm{x}) + \epsilon(\bm{x}),
	\end{equation}
	wherein $ f_h(\bm{x})$ is function of $\bm{x}$, $\beta_h$ are unknown coefficients to be estimated, $\epsilon(\bm{x})$ is the error term, which is a random process with zero mean and covariance
	\begin{equation}
		\text{Cov}\left[\bm{x}_{i},\bm{x}_{j}\right] = \sigma^2 \operatorname{Corr}\left[\bm{x}_{i},\bm{x}_{j}\right], 
	\end{equation}
	where $\bm{x}_{i}$ and $\bm{x}_{j}$ are two sets of variables, $\sigma$ is the process variance and $\operatorname{Corr}\left[\bm{x}_{i},\bm{x}_{j}\right]$ represents the correlation. 
	Equation (\ref{Eq:Y_x}) can be regarded as the Bayesian prior on the true response function, in which the right hand side resembles the form of linear regression model, except that the errors of different $\bm{x}$ are correlated rather then independent with each other. We consider the common and simple case where the stochastic process is stationary, and the correlation takes the specific form of\cite{jeromesacksDesignAnalysisComputer1989} 
	\begin{equation}\label{Eq:Cor_1}
		\operatorname{Corr}\left[\bm{x}_{i},\bm{x}_{j}\right]=\exp \left[-d\left(\bm{x}_{i}, \bm{x}_{j}\right)\right],
	\end{equation}
	with
	\begin{equation}\label{Eq:Cor_2}
		d\left(\bm{x}_{i}, \bm{x}_{j}\right)=\sum_{h=1}^k \alpha_h\left|x_{i,h}-x_{j,h}\right|^{p_h} \quad\left(\alpha_h \geq 0, p_h \in[1,2]\right), 
	\end{equation}
	where $\alpha_h$ and $p_h$ are parameters that need to be determined. This specific form corresponding to a product of Ornstein–Uhlenbeck process at $p_h = 1$, which is widely used in physical science models. Furthermore, with the correlation form of Equation (\ref{Eq:Cor_1}) and (\ref{Eq:Cor_2}), the regression parts $\sum_{h}\beta_h f_h(\bm{x})$ in the stochastic process model can be simply replaced by a constant $\mu$ without undermining the predictive performance\cite{jonesEfficientGlobalOptimization1998}. Hereafter we use the simplified stochastic process model 
	\begin{equation}\label{Eq:simp Y_x}
		Y(\bm{x}) =\mu+ \epsilon(\bm{x}).
	\end{equation}

 	A typical methods for analyzing such stationary Gaussian process model is the Kriging method\cite{matheronPrinciplesGeostatistics1963}, which is flexible and robust for global estimation\cite{ngBayesianKrigingAnalysis2012} and is widely applied in fields of spatial statistics\cite{matheronPrinciplesGeostatistics1963,steinInterpolationSpatialData1999}, design of computer experiments \cite{jeromesacksDesignAnalysisComputer1989,martinUseKrigingModels2005a} and Bayesian optimization \cite{jonesEfficientGlobalOptimization1998}. Besides the ordinary Kriging method, various analysis methods based on Kriging method as well as Bayesian approach are also well studied for Gaussian process with non-stationary correlation and other kinds of stochastic process\cite{currinBayesianApproachDesign1988, currinBayesianPredictionDeterministic1991, morrisBayesianDesignAnalysis1993,ngBayesianKrigingAnalysis2012,kleijnenKrigingMetamodelingSimulation2009}. Here we adopt the best linear unbiased predictor (BLUP) of Equation (\ref{Eq:simp Y_x}) given by the Kriging method to form the estimation model. Supposing the response function has been evaluated at $n$ samples $\bm{s} = \{\bm{s}_1; \bm{s}_2; \cdots; \cdots \bm{s}_n\}$, and the corresponding function values are $y(\bm{s}) = \{y(\bm{s}_1),y(\bm{s}_2),\cdots y(\bm{s}_n)\}^{\prime}$. The BLUP of $y(\bm{x})$ can be represented as \cite{jonesEfficientGlobalOptimization1998}
	\begin{equation}\label{Eq:simp BLUP}
		\hat{y}(\bm{x})=\hat{\mu}+r^{\prime}(\bm{x}) \mathbf{R}^{-1}\left(\bm{y}(\bm{s})-\mathbbm{1} \hat{\mu}\right),
	\end{equation}
 	where $\mathbf{R}$ is $n\times n$ correlation matrix with entries $R_{i,j} = \text{Corr}\left[\bm{s}_i, \bm{s}_j\right]$, $r(\bm{x})=\left[R\left(\bm{s}_1, \bm{x}\right), \cdots, R\left(\bm{s}_n, \bm{x}\right)\right]^{\prime} $ is the vector of correlation between the errors at sample and untried input $\bm{x}$, with $R(\bm{s}_1,\bm{x}) = \text{Corr}\left[(\bm{s}_1,\bm{x}\right]$, and $\hat{\mu} = \left(\mathbbm{1}^{\prime} \mathbf{R}^{-1} \mathbbm{1}\right)^{-1} \mathbbm{1}^{\prime} \mathbf{R}^{-1} \bm{y}(\bm{s})$ is the generalized least-squares estimation of $\mu$.

	Selecting the parameters in Equation (\ref{Eq:Cor_2}) properly we can expect a predict function that describing the true function $y(\bm{x})$ well using only the limited number of sample points $\bm{s}$. One useful method to determine these parameters is the maximum likelihood estimation (MLE). For the Gaussian process we take here, the likelihood function of the samples are \cite{jonesEfficientGlobalOptimization1998}
	\begin{equation}\label{eq:likelihood function}
		\frac{1}{(2 \pi)^{n / 2}\left(\sigma^2\right)^{n / 2}|\mathbf{R}|^{\frac{1}{2}}} \exp \left[-\frac{(\bm{y}(\bm{s})-\mathbbm{1} \mu)^{\prime} \mathbf{R}^{-1}(\bm{y}(\bm{s})-\mathbbm{1} \mu)}{2 \sigma^2}\right],
	\end{equation}
	where $|\mathbf{R}|$ represents the determinant of $\mathbf{R}$. The analytical solutions of $\mu$ and $\sigma$ maximizing Equation (\ref{eq:likelihood function}) are
	\begin{equation}
		\hat{\mu} = \left(\mathbbm{1}^{\prime} \mathbf{R}^{-1} \mathbbm{1}\right)^{-1} \mathbbm{1}^{\prime} \mathbf{R}^{-1} \bm{y}(\bm{s}),
	\end{equation}
	which is exactly the generalized least-squares estimation of $\mu$, and
	\begin{equation}
		\hat{\sigma}^2=\frac{(\bm{y}(\bm{s})-\mathbbm{1} \hat{\mu})^{\prime} \mathbf{R}^{-1}(\bm{y}(\bm{s})-\mathbbm{1} \hat{\mu})}{n}.
	\end{equation}
	Such we only need to find the values of $\alpha_h$ and $p_h$ that give the maximum value of Equation (\ref{eq:likelihood function}), which can be completed by any optimization method at hand.
	
	Corresponding to the specific model of two-level NV electron spin system, once the control field $g(t)$ and control time $T$ are fixed, the target function $f(\delta,\kappa)$ is the deterministic response function of two-dimension variable $\bm{x} = \{ \delta, \kappa\}'$, and can be approximately estimated by the predict function $\hat{f} (\bm{x})$ which takes the same form of $\hat{y} (\bm{x})$ in Equation (\ref{Eq:simp BLUP}). Specifically, we write the Hamiltonian in the interaction picture with respect to $\frac{\omega_0}{2}\sigma_z$ as
	\begin{equation}\label{Eq:H_delta_kappa}
		H_{\delta,\kappa}(t) = \frac{\delta}{2}\sigma_z + \kappa H_c(t),
	\end{equation}
	where $H_c(t)$ is the time-depended Hamiltonian. Neglecting counter-rotating terms in a rotating-wave approximation, $H_c(t)$ can be expressed as
	\begin{equation}
		H_c(t)=\Omega_x(t) \sigma_x + \Omega_y(t) \sigma_y,
	\end{equation}
	where $\Omega_x(t)$ and $\Omega_y(t)$ take the form
	\begin{equation}
		\Omega^{\mathrm{SFB}}_x(t)=\sum_{j=1}^{N_{\text{D}}} \frac{a_j}{2} \cos \left(\omega_j t+\phi_j\right)\cos \left(\varphi_j\right),
	\end{equation}
	\begin{equation}
		\Omega^{\mathrm{SFB}}_y(t)=\sum_{j=1}^{N_{\text{D}}} \frac{a_j}{2} \cos \left(\omega_j t+\phi_j\right)\sin \left(\varphi_j\right)
	\end{equation}
	for SFB method, and the form
	\begin{equation}
		\Omega^{\mathrm{PM}}_x(t)=\sum_{j=1}^{N_{\text{D}}} \frac{a_j}{2}\cos \left[\frac{b_j}{v_j} \sin \left(v_j t\right)\right],
	\end{equation}
	\begin{equation}
		\Omega^{\mathrm{PM}}_y(t)=\sum_{j=1}^{N_{\text{D}}} \frac{a_j}{2}\sin	\left[\frac{b_j}{v_j} \sin \left(v_j t\right)\right]
	\end{equation}
	for PM method. The target response function can be expressed as
	\begin{equation}
		f(\delta,\kappa) = \left|\left\langle\psi_g \mid U_{\delta,\kappa} \psi(0)\right\rangle\right|^2,
	\end{equation}
	where
	\begin{equation}\label{Eq:U_delta_kappa}
		U_{\delta,\kappa}=\mathcal{T} \exp \left[-i \int_0^T H\left(\delta,\kappa, t\right) d t\right]
	\end{equation}
	is the evolution operator and $\mathcal{T}$ is the time-order operator.  
	
	So far we have established the Bayesian based estimation model of the final state fidelity under certain frequency detuning $\delta$ and amplitude bias ratio $\kappa$. With this usable tool, we could go further to construct a high efficiency optimization method to explore the optimal control field of ensemble system.
	
	\subsection{The hybrid optimization method}
	\begin{figure*}[] 
		\centering 
		\includegraphics[width = 15cm]{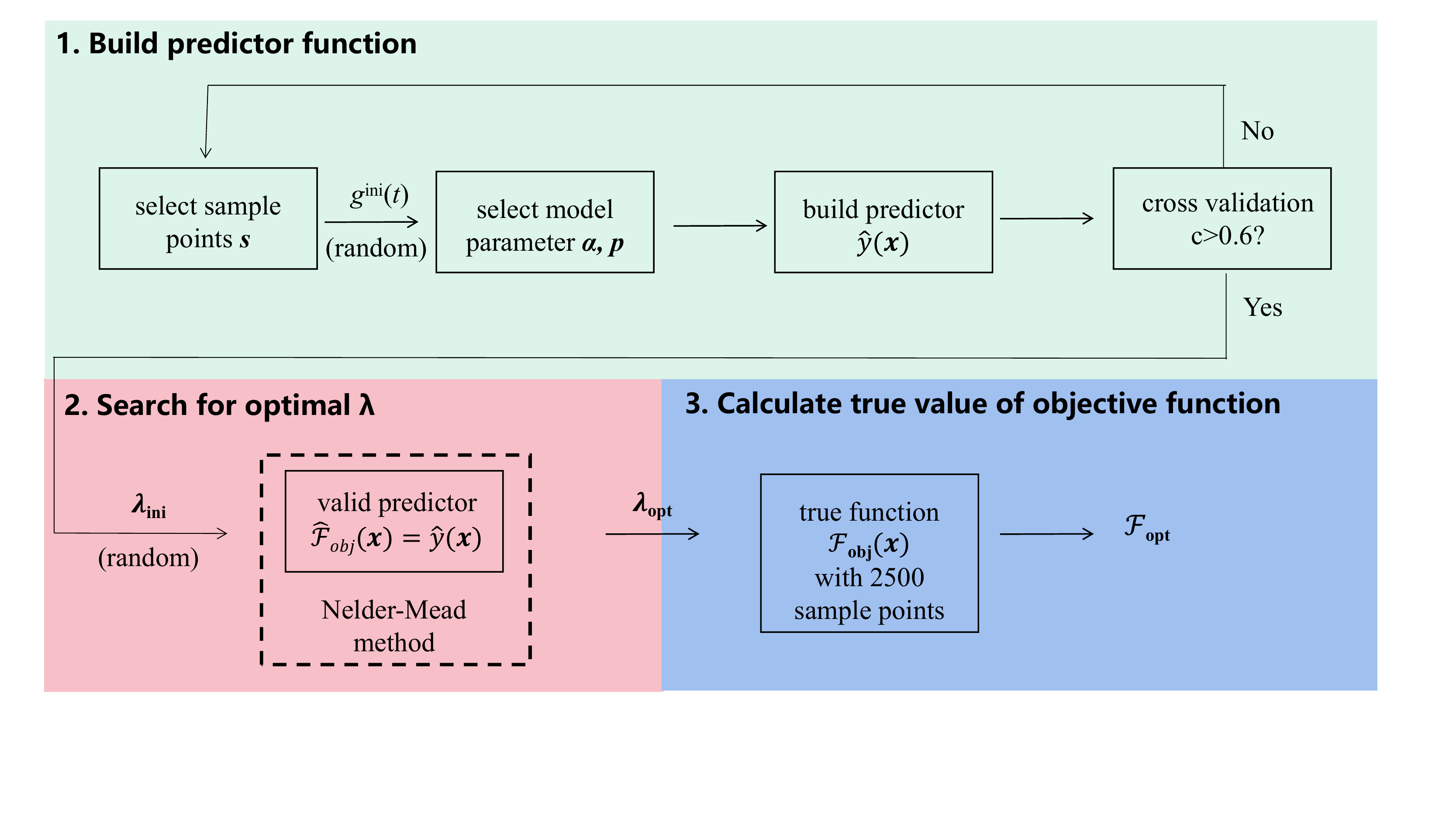}
		\caption{Process schematic of Bayes-based optimization method.}
		\label{Fig_2}
	\end{figure*}   
	Now we introduce the hybrid B-PM optimization method for solving the problem described by Equation (\ref{Eq:optim_model}). As represented in Figure~\ref{Fig_2}, the whole optimization process can be divided into three parts, initially construct the available predict function, then search for the optimal parameter $\bm{\lambda}$ using the predict function as the objective function, lastly calculate the true value of objective function in Equation (\ref{Eq:Fobj}) with $M\times N = 2500$ sample points. The first segment is the essential part of this method, and the following strategies are adopted to ensure the convergence and improve the optimization results.
	
	\begin{enumerate}
		\item	Sample position: In the sample selection step, instead of completely randomly picking the sample points, we add random bias on even-distributed coordinate values to give the randomly yet uniformly distributed sample position. This choice improves the estimation performance especially in cases with small sample number.
		\item	Model parameter: The model parameter $\alpha$ and $p$ are selected based on a random initial filed $g^{\text{ini}}(t)$, and is fixed during one optimization process. This tactic ensures the monodromy of the objective function and the convergence of the subsequent search process, reduces the computation time spend on building the predict function, and doesn't damage the estimation accuracy. Detailed verification are showed in Figure~\ref{Fig_4} of next section.
		\item	Cross valiation: After constructing a predict function, an cross valiation is applied to eliminate the low-performing ones. Each function value of the $n$ sample points $\bm{y}(\bm{s})$ is successively regarded as unknown quantity and is predicted based on the model parameter $\alpha, p$ and other $n-1$ samples, given a corresponding predict vector $\bm{y}_{\text{pred}}(\bm{s})$. Then a linear fitting is performed on those points located at $\left(\bm{y}(\bm{s}),\bm{y}_{\text{pred}}(\bm{s})\right)$ and we use the slope $p_{\text{fit}}$ as a criterion. In principle, a $p_{\text{fit}}$ far away from $1$ indicates a bad performance of the predictor, and by examining the fitted slop of bad predict models occurred in the optimization process we found bad models corresponding to small values of $p_{\text{fit}}$. Therefore we set $p_{\text{fit}}>0.6$ as the threshold. If this condition is fulfilled, the predict function would be used as the objective function in the following direct search process. Otherwise a new model need to be build from the very beginning.
	\end{enumerate}

	After the valid predictor is constructed, it is used as the objective function of the following direct search process, where we using the phase-modulated to further improve the efficiency.  The initial parameter $\bm{\lambda}_{\text{ini}}$ is taken in the range $a^{\text{ini}}_j \in [0, \Omega_{\text{max}}]$ and $b^{\text{ini}}_j, \nu^{\text{ini}}_j \in [0, 2\pi/T]$, with $\Omega_{\text{max}}$ being the maximum field amplitude and $T$ the evolution time. The SFB method is also investigated for comparison, with the initial parameter $a^{\text{ini}}_j \in [0, \Omega_{\text{max}}]$, $\omega^{\text{ini}}_j \in [0, 2\pi/T]$ and $\phi^{\text{ini}}_j, \varphi^{\text{ini}}_j \in [0, 2\pi]$. For both methods, we take $\Omega_{\text{max}}=2\pi\times10$ MHz and $T = 100$ ns, and the total field amplitude is constrained as $\sqrt{\Omega_x^2(t)+\Omega_y^2(t)}\leq \Omega_{\text{max}}$, the frequency parameters $b_j, \nu_j$ and $\omega_j$ are constrained in range $[0, 5\times 2\pi/T]$, and the phase parameters $\phi_j$ and $\varphi_j$ are constrained in range $[0, 2\pi]$.
	
	In the last step, given the optimization parameter $\bm{\lambda}_\text{opt}$, the average fidelity in Equation(\ref{Eq:Fobj}) is calculated based on $2500$ uniformly spaced sample points in region $\delta \in 2\pi \times [-10,10]$ MHz and $\kappa\in [0.5,1.5]$, with $M=N=50$.
	
	\subsection{AC magnetometry}
	Magnetometry is one of the most promising field that NV center can play an important part with its outstanding characters such as fine biological compatibility, normal pressure and temperature operating conditions, and relatively long spin lifetimes. Pulsed and continuous dynamical decoupling (DD) has been applied in coherent control and magnetic sensing of NV center. In practice, large inhomogeneous broadening of NV center ensemble will decrease the controllability of DD sequence, and ultimately hinder the sensitivity in magnetometry. Adiabatic strategy has been applied to tackle such problem and achieve good results. When the frequency of signal is not low enough to fulfill the adiabatic condition, finding optimal field with limited pulse length to achieve certain gate operation become a key approach to improve the sensitivity. Here we use the B-PM method to optimize the control field and gives robust Pauli-X and Pauli-Y gate under inhomogeneous broadening, which composes the XY-8 DD sequence in the AC magnetometry process, see the sketch in Figure~\ref{Fig_7} (a).

	During the pulse period, the control Hamiltonian $H_\text{c}$ in Equation (\ref{Eq:H_delta_kappa}) takes the form of
	\begin{equation}
		H_{\mathrm{PM}}^{\mathrm{X}}(t)=\Omega_1(t) \sigma_x+\Omega_2(t) \sigma_y
	\end{equation}
	for Pauli-X and
	\begin{equation}
		H_{\mathrm{PM}}^{\mathrm{Y}}(t)=\Omega_1(t) \sigma_y-\Omega_2(t) \sigma_x
	\end{equation}
	for Pauli-Y gate, with
	\begin{equation}
		\Omega_1(t)=\sum_{j=1}^{N_{\mathrm{D}}} \frac{a_j}{2}\cos \left[\frac{b_j}{v_j} \sin \left(v_j t\right)\right], \quad \Omega_2(t)=\sum_{j=1}^{N_{\mathrm{D}}} \frac{a_j}{2}\sin \left[\frac{b_j}{v_j} \sin \left(v_j t\right)\right].
	\end{equation}
	\
	For gate optimization the objective function takes the form
	\begin{equation}
		\mathcal{F}_{\mathrm{obj}}=\mathcal{N} \sum_{k=1}^M \sum_{j=1}^N p\left(\delta_k\right) p\left(\kappa_j\right) f_g\left(\delta_k, \kappa_j\right),
	\end{equation}
	where
	\begin{equation}
		f_g(\delta,\kappa)=\frac{1}{2}+\frac{1}{3} \sum_{\epsilon=x, y, z} \operatorname{Tr}\left(U_{\mathrm{Tar}} \frac{\sigma_\epsilon}{2} U_{\mathrm{Tar}}^{\dagger} U_{\delta,\kappa} \frac{\sigma_\epsilon}{2} U_{\delta,\kappa}^{\dagger}\right),
	\end{equation}
	is the gate fidelity of fixed $\delta$ and $\kappa$, $U_{\mathrm{Tar}}$ is the target quantum gate, $U_{\delta,\kappa}$ is the actual operator,takes the same form as in Equation (\ref{Eq:U_delta_kappa}), $\mathcal{N}, p\left(\delta_k\right)$ and $p\left(\kappa_k\right)$ take the same form and value as in Equation (\ref{Eq:Fobj}). We apply the B-PM method to search for the robust control field with the amplitude constraint $\sqrt{\Omega_1^2(t)+\Omega_2^2(t)}\leqslant \Omega_{\max }$, and obtain an optimized shaped pulse which constructs the XY-8 DD sequence in the AC magnetometry strategy.
	
	As is showed in Figure~\ref{Fig_7} (a), the AC magnetic signal to be measured oscillates at frequency $\omega_{\mathrm{s}}$, and the control pulses is applied on time nodes where the AC signal changes its direction. Due to the power limitation of the microwave field, each pulse possesses a finite pulse length $T_{\text{pulse}}$, and its value follows the relationship
	$\omega_s=\pi /\left(T_{\text {pulse }}+\tau_{\text{pulse}}\right)$, where $\tau_{\text{pulse}}$ is the time separation between two adjacent pulses.
	The amplitude of the AC magnetic signal $g_{\text{ac}}$ can be read out from the Ramsey oscillation of the NV center sensor. At the beginning of the measurement, the NV centers in ensemble are initialized into state $\ket{0}$, flipped by a $\pi/2$ pulse and begin to rotate around $\sigma_z$ axis. After a time period $t$, one $3\pi/2$ pulse is applied to the sensors and their population on $\ket{0}$ is measured. In ideal condition with instantaneous gate pulses, the population can be expressed as $P_0=\left(1+\cos (2\chi(t))\xi(t) \right)/2$, with the phase term
	$\chi(t) \equiv \int_0^t g\left|\cos \left(\omega_{\mathrm{s}} t^{\prime}\right)\right| d t^{\prime}$, and the exponential decay term $\xi(t)$ induced by the inhomogeneous broadening and dynamic noise. At coherent time $t = T_2$, the decay term goes down to $1/e$. In practice, the finite pulse length leads to a deviation of the real phase term from $\chi(t)$.

	Specifically, the total Hamiltonian in the rotating wave approximation is
	\begin{equation}\label{Eq:H_gate}
		H_{\delta, \kappa}(t)=\frac{\delta+\delta_d(t)}{2} \sigma_z+g_{\text{ac}}\cos(\omega_{s}t) \sigma_z + \kappa H_c,
	\end{equation}
	where a time-dependent dynamic noise term $\delta(t)$ is considered, formed by the Ornstein-Uhlenbeck process 
	\begin{equation}
		\delta_d(t+\Delta t)=\delta_d(t) e^{-\Delta t / \tau}+\left[\frac{c \tau}{2}\left(1-e^{-2 \Delta t / \tau}\right)\right]^{1 / 2} n,
	\end{equation}
	where $\tau$ and $c$ is the correlation time and the diffusion constant of the noise respectively, valued as $\tau = 20 \ \mu\mathrm{s}$ and $\sqrt{c \tau / 2}=2 \pi \times 50 \mathrm{KHz}$\cite{genovEfficientRobustSignal2020a}, and $n\sim \mathcal{N}(0,1)$. The controlled Hamiltonian $H_
	\text{c}$ only take a non-zero value during the pulse duration $T_{\text{pulse}}$. We take the frequency of the signal to be $\omega_s=\pi /\left(T_{\text {pulse }}+\tau_{\text{pulse}}\right)=2.5 \pi \ \mathrm{MHz}$, where $\tau_{\text{pulse}}$ is the time separation between two adjacent pulses. We take the maximum limit of field amplitude as $\Omega_{\max } = 2\pi\times 10 \ \mathrm{MHz}$, so for rectangular pulse, $T_{\text {pulse }} = 50 \ \mathrm{ns}$, during which $H_c = \Omega_{\max }\sigma_x$ for Pauli-X gate and  $H_c = \Omega_{\max }\sigma_y$ for Pauli-Y gate, and the time separation is $\tau_{\text{pulse}} = 350 \ \mathrm{ns}$. For optimization pulses we set  $T_{\text {pulse }} = 100 \ \mathrm{ns}$, $\tau_{\text{pulse}} = 300 \ \mathrm{ns}$. 
	
	%%%%%%%%%%%%%%%%%%%%%%%%%%%%%%%%%%%%%%%%%%
	\section{Results}
	\subsection{Feasibility of the estimation model}
	\begin{figure*}[]
		\centering 
		\includegraphics[width = 12cm]{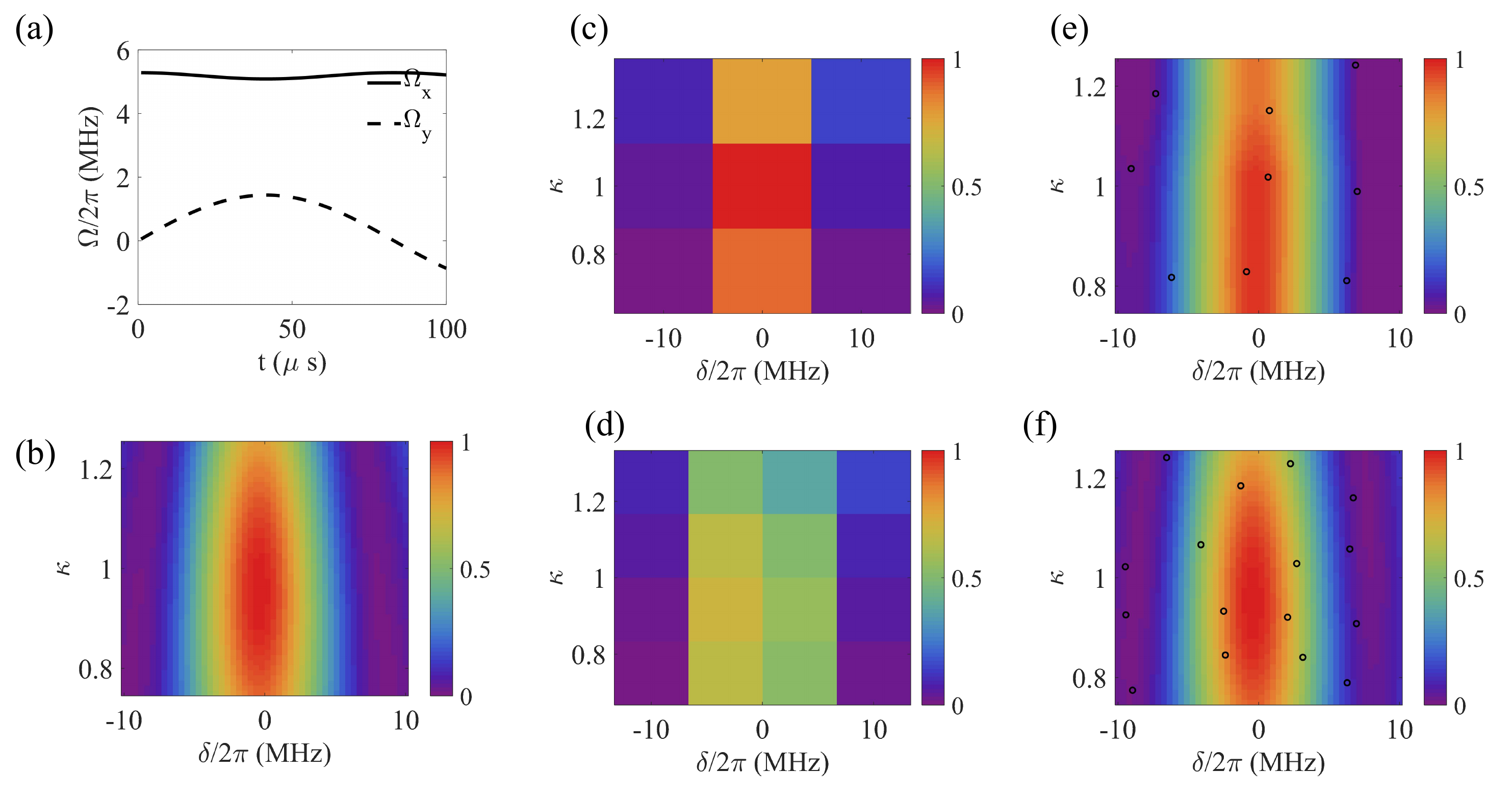}
		\caption{(\textbf{a}) The shape of control field in the interaction picture, $\Omega^{\mathrm{PM}}_x(t)=\sum_{j=1}^{N_{\text{D}}} \frac{a_j}{2}\cos \left[\frac{b_j}{v_j} \sin \left(v_j t\right)\right]$ and $\Omega^{\mathrm{PM}}_y(t)=\sum_{j=1}^{N_{\text{D}}} \frac{a_j}{2}\sin	\left[\frac{b_j}{v_j} \sin \left(v_j t\right)\right]$, with the randomly taken parameters $a = 0.0332$, $b = 0.0104$ and $\nu=0.0378$. (\textbf{b}) (\textbf{c}) Values of true function with sample number $N_s = 9$. (\textbf{d}) Values of true function with sample number $N_s = 16$. (\textbf{e}) Values of estimation function with sample number $n = 9$ and estimation points number $N_e = 2500$. (\textbf{f}) Values of estimation function with sample number $n = 16$ and estimation points number $N_e = 2500$. Black circle in (\textbf{e}) and (\textbf{f}) represent the locations of the sample points.}
		\label{Fig_3}
	\end{figure*}
	
	\begin{figure}[]
		\centering   
		\includegraphics[width = 8cm]{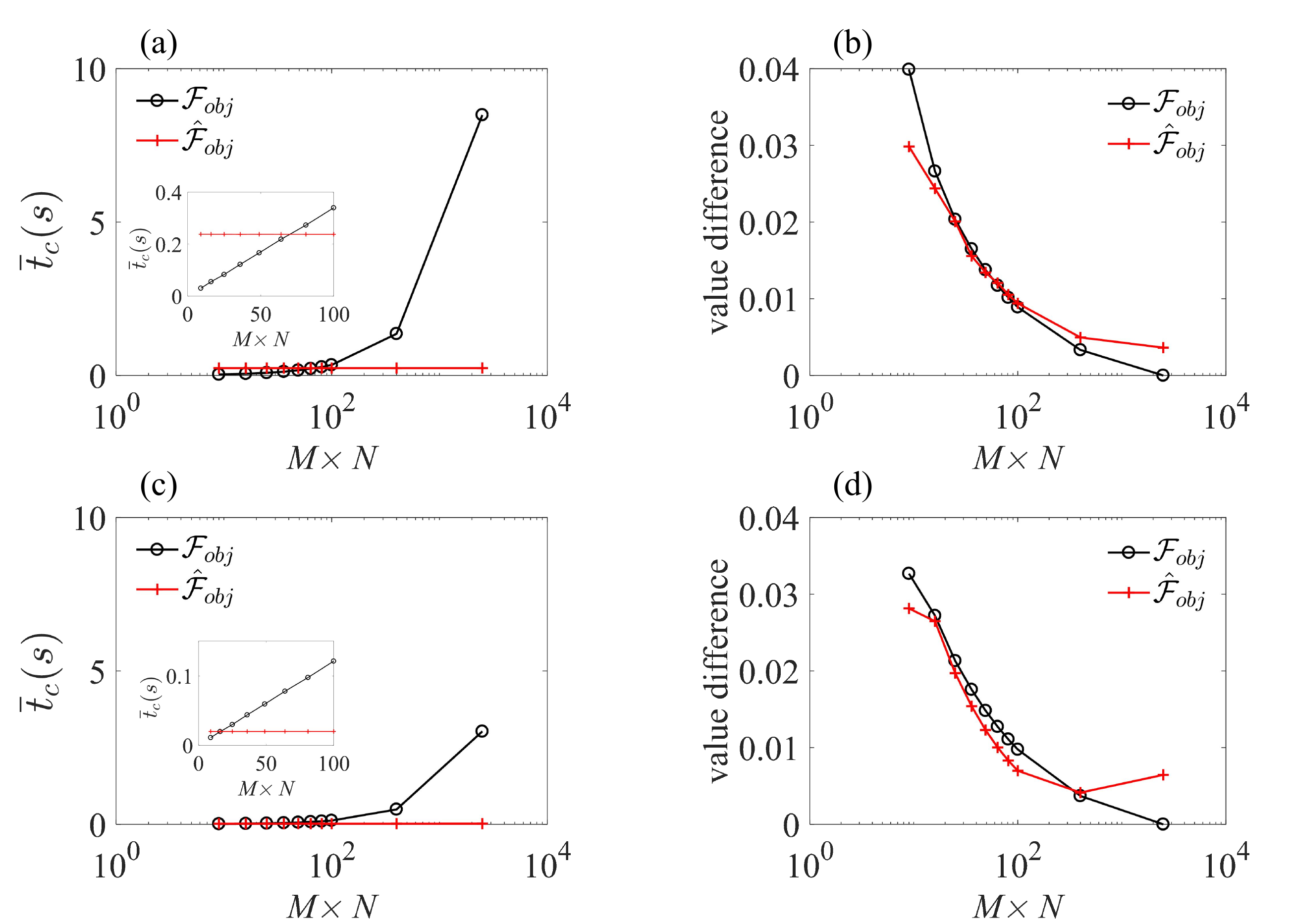}
		\caption{\textbf{(a)} Average computation time of the objective function as a function of calculating sample number $M\times N$. $100$ processes under random control fields are calculated, and the estimation model parameter is updated  in each process when computing $\hat{\mathcal{F}}_\text{obj}$. \textbf{Insert}: The zoom-in graph for $M\times N \le 100$. \textbf{(b)} Average function value deviation as a function of calculating sample number $M\times N$. Based on the same data with \textbf{(a)}. The deviation is calculated by subtracting the values of $f(\delta,\kappa)$ with $M\times N = 2500$, i.e., $|\mathcal{F}_\text{obj}-\mathcal{F}_{\text{obj}(M\times N=2500)}|$ for true value based process and  $|\hat{\mathcal{F}}_\text{obj}-\mathcal{F}_{\text{obj}(M\times N=2500)}|$ for predictor based process. \textbf{(c)} Average computation time of the objective function as a function of calculating sample number $M\times N$, while the estimation model parameter is fixed in all the $100$ random processes.  \textbf{Insert}: The zoom-in graph for $M\times N \le 100$. \textbf{(d)} Average function value deviation as a function of calculating sample number $M\times N$. Based on the same data with \textbf{(c)}. The control fields take the form $g(t)=a\cos \left[\omega_0 t+\frac{b}{\nu} \sin \left(\nu t\right)\right]$, with the evolution time $T = 100$ ns, and $a, b$ and $\nu$ randomly taken from the range $a \in [0, 10\times 2\pi]$ MHz, $b \in [0, 2\pi/T]$ and $\nu \in [0, 2\pi/T]$. The sample number used in the estimation model is $n = 16$.}
		\label{Fig_4} 
	\end{figure}
	
	We first demonstrate the feasibility of the estimation model based on its estimation accuracy and time efficiency. Figure~\ref{Fig_3} shows its performance on predicting the fidelity of final state under certain frequency detuning $\delta$ and amplitude drift factor $\kappa$. For a fixed control field showed in Figure~\ref{Fig_3} (a), the value of $f(\delta,\kappa)$ varies with $\delta$ and $\kappa$, which we show by a 2-D color distribution image in Figure~\ref{Fig_3} (b), where $50\times50$ sample points are taken to ensure it's resolution. If the available sample number decreases to $9$ and $16$ respectively, the resolution ratio drops clearly, as shown in Figure~\ref{Fig_3} (c) and (d). However, based on the same number of sample points, the image depicted by $2500$ prediction values $\hat{f} (\bm{x})$ in Figure~\ref{Fig_3} (e) and (f) shows an obviously improvement in the resolution ratio and the accuracy compared to their counterpart in Figure~\ref{Fig_3} (c) and (d) respectively. By applying the estimation method, more information about how the target function distributed in the parameter space is extracted based on the knowledge provided by the sample points. 
	
	Another concern is how much time would this prediction process takes, or if it has advantage on computing time and cost. Here we use the average computing time of $100$ prediction process to give a general description. The control field in each process takes the form of $g(t)=a\cos \left[\omega_0 t+\frac{b}{\nu} \sin \left(\nu t\right)\right]$, with evolution time $T= 100$ ns, and the parameter $a, b$ and $\nu$ are randomly taken from the range $a \in [0, 10\times 2\pi]$ MHz, $b \in [0, 2\pi/T]$ and $\nu \in [0, 2\pi/T]$. Each prediction process using $16$ sample point, and the mission is to calculate the average fidelity (see Equation (\ref{Eq:Fobj}) ) in range $\delta \in 2\pi \times [-10,10]$ MHz and $\kappa\in [0.5,1.5]$. We represent the estimated fidelity as
	\begin{equation}\label{Eq:F_est}
		\hat{\mathcal{F}}_{\mathrm{obj}}=\mathcal{N} \sum_{k=1}^M \sum_{j=1}^N p\left(\delta_k\right) p\left(\kappa_j\right) \hat{f}\left(\delta_k, \kappa_j\right),
	\end{equation}
	where the formation and values of $\mathcal{N}, p\left(\delta_k\right)$ and $p\left(\kappa_k\right)$ are the same with Equation (\ref{Eq:Fobj}), and $M\times N$ represents the number of different predict points used in the calculation. Two strategies are respectively applied: in Figure~\ref{Fig_4} (a), the estimation model are updated in each trials, while in Figure~\ref{Fig_4} (c) a fixed estimation model serves in all the $100$ trials. To make a comparison, we also record the results of calculating the true $\mathcal{F}_{\mathrm{obj}}$ by $M\times N$ sample points. It's natural that for the true function based objective function $\mathcal{F}_{\mathrm{obj}}$, the computation time will be approximately proportional to $M\times N$, the total calling number of function $f(\delta,\kappa)$. In contrast, the processing time of the predict function based objective function remains stable as $M\times N$ increases, bringing a distinct time retrench as $M\times N$ rise up to $2500$. Notice here for $M\times N\le64$, the computation time of $\hat{\mathcal{F}}_\text{obj}$ is longer than those of $\mathcal{F}_\text{obj}$, because values of $16$ sample points need to calculated, and an extra amount of time need to be spent on selecting the parameter that construct the estimation model. Figure~\ref{Fig_4} (c) shows once the estimation model is built, process of calculating the value of predict function become rapid and its corresponding time consuming can be neglected, so the total computation time is decided by the number of true sample points, i.e. $M\times N = 16$.
	
	Figure~\ref{Fig_4} (b) and (d) shows the estimation accuracy of $\hat{\mathcal{F}}_\text{obj}$ and $\mathcal{F}_\text{obj}$ with different calculated sample number $M\times N$. Figure~\ref{Fig_4} (b) uses the same data with Figure~\ref{Fig_4} (a), Figure~\ref{Fig_4} (d) uses the same data with Figure~\ref{Fig_4} (c). We use the value of $\mathcal{F}_\text{obj}$ with $M \times N=  2500$ as the benchmark to calibrate the accuracy, denoted as $\mathcal{F}_{\text{obj}(M\times N =2500)}$. The estimation accuracy is expressed as the value deviation from $\mathcal{F}_{\text{obj}(M\times N =2500)}$, that is  $|\hat{\mathcal{F}}_\text{obj}-\mathcal{F}_{\text{obj}(M\times N =2500)}|$ for estimation objective function method and $|\mathcal{F}_\text{obj}-\mathcal{F}_{\text{obj}(M=50)}|$ for true objective function method. In general, the deviation decreases as $M\times N$ increase, and the estimation deviation values of Figure~\ref{Fig_4} (b) and (d) are in the same scope, shows that the fixed estimation model strategy doesn't damage the estimation accuracy.
	
	\subsection{Optimization efficiency of the B-PM method}  
	\begin{figure}[] 
		\centering   
		\includegraphics[width = 8cm]{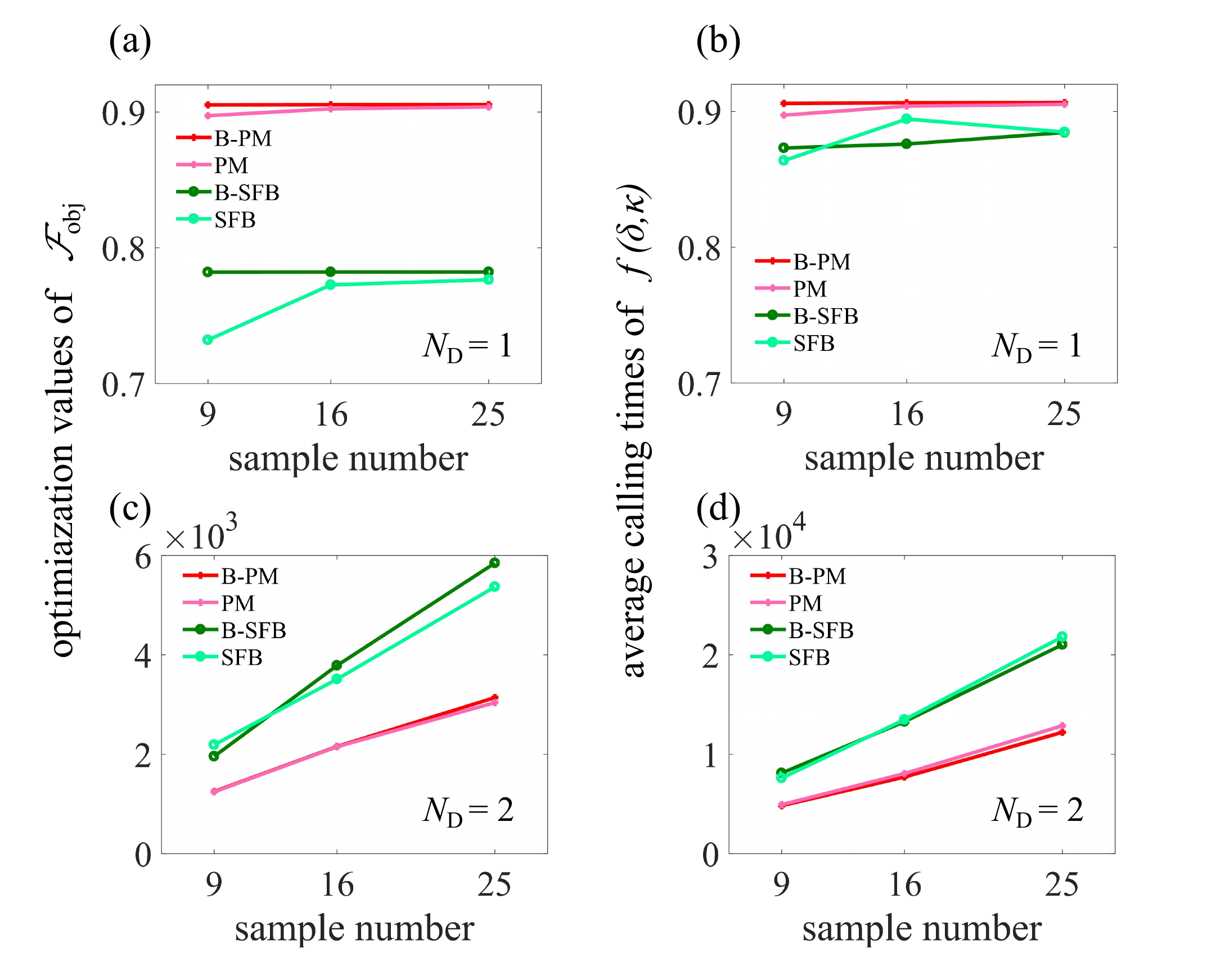}
		\caption{\textbf{(a)} The optimized fidelity of B-PM, PM, B-SFB and SFB methods with parameter sets number $N_{\text{D}} = 1$. \textbf{(b)} The optimized fidelity of B-PM, PM, B-SFB and SFB methods with parameter sets number $N_{\text{D}} = 2$. \textbf{(c, d)} Average calling times of function $f(\delta, \kappa)$ in Equation (\ref{Eq:single_f})) that gives the results in \textbf{(a)}  and \textbf{(b)}  respectively. All results are based on $100$ trials with random initial parameters. The total evolution time is taken as $T = 100$ ns, the maximal field amplitude is bounded as $\max |g(t)| \leqslant \Omega_{\max }= 2\pi \times 10$ MHz.}
		\label{Fig_5}
	\end{figure}

	The results in Figure~\ref{Fig_4} has demonstrated using fixed estimation model, the total computation time is decided by the calling time of the function $f(\delta,\kappa)$. Figure~\ref{Fig_5} shows a comparison between B-PM method, PM method, Bayesian estimation SFB (B-SFB) method and SFB method in terms of final value of $\mathcal{F}_{\text {obj }}$ and calling times of function $f(\delta,\kappa)$ during the search process. Results with $N_{\text{D}} = 1$ are showed in Figure~\ref{Fig_5} (a) and (c), results with $N_{\text{D}} = 1$ are showed in Figure~\ref{Fig_5} (b) and (d). Each results are based on $100$ trials with random initial parameters. On the whole, the B-PM method gives result of $\mathcal{F}_{\mathrm{obj}}= 0.905$ with average calling times $1252$, while the SFB method gives result of $\mathcal{F}_{\mathrm{obj}}= 0.894$ with average calling times $13479$, such the B-PM reaches a improved average fidelity using only $9.3\%$ computation resource compared with SFB method. 
	
	\begin{figure*}[] 
		\centering   
		\includegraphics[width = 12cm]{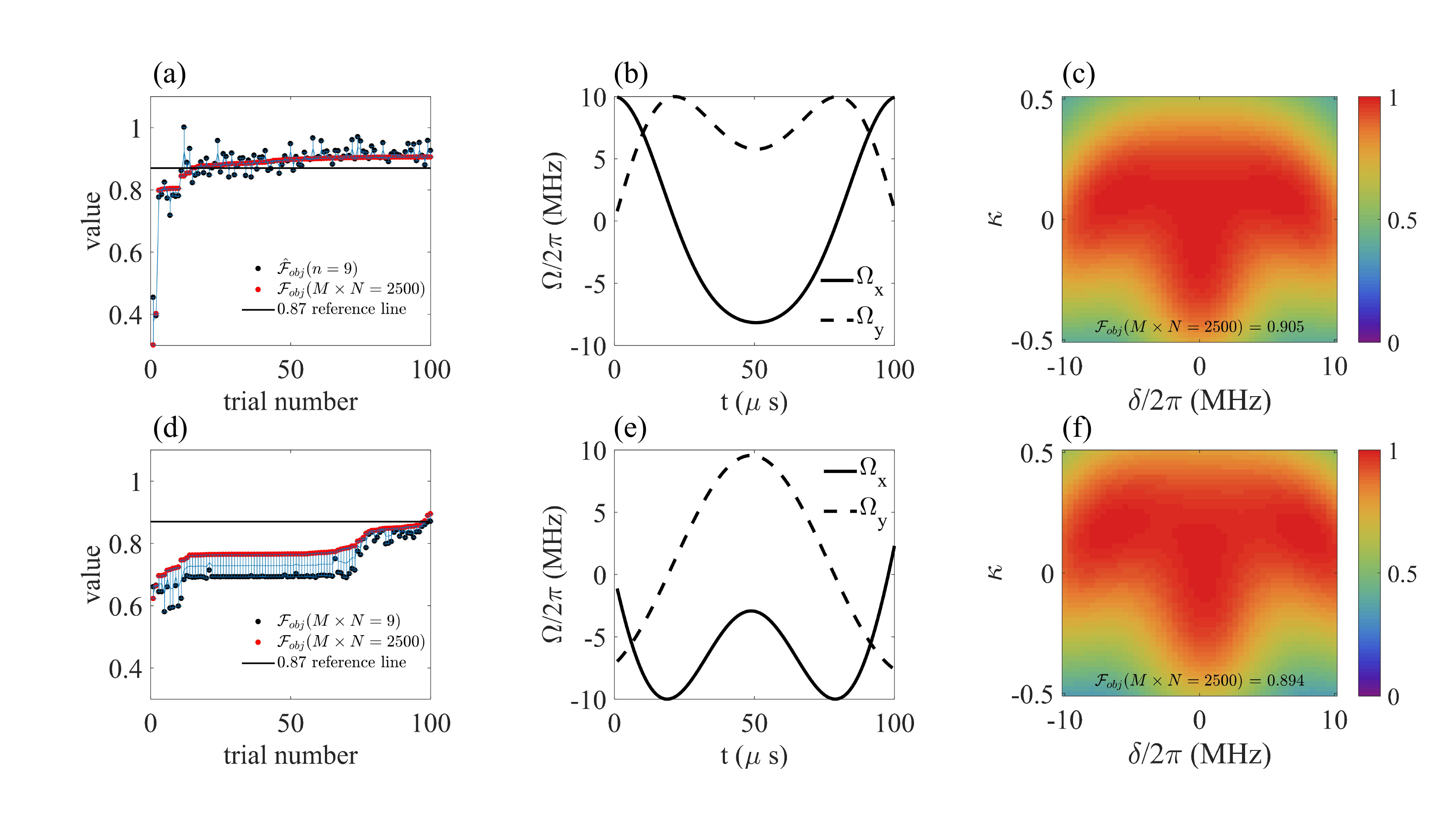}
		\caption{\textbf{(a-c)} Optimization results given by B-PM method with $N_\text{D} = 1$, $n = 9$. \textbf{(d-f)} Optimization results given by SFB method with $N_\text{D} = 2$, $M\times N = 16$. \textbf{(a, d)} Optimization values of objective function of $100$ trials with random initial parameters $\bm{\lambda}$. \textbf{(b, e)} The shape of optimized control field in the interaction picture. \textbf{(c, f)} The fidelity distribution in region  $\delta \in 2\pi \times [-10,10]$ MHz and $\kappa\in [0.5,1.5]$.}
		\label{Fig_6}
	\end{figure*}
	
	Figure~\ref{Fig_6} shows the detail optimization results of B-PM  method with $N_{\text{D}} = 1, n = 9 $ and SFB method with $N_{\text{D}} = 2, M\times N = 16$. Figure~\ref{Fig_6} (a) and (d) shows the final objective function values of $100$ trials with random initial parameters $\bm{\lambda}$, in which we mark the value used in the optimization process by black points, and their corresponding more accurate value calculated by $M\times N = 2500$ true sample points by red points. As is showed in Figure~\ref{Fig_6} (a), among $100$ trials of B-PM method, $42$ results give $\mathcal{F}_{\mathrm{obj}}\geq0.9$ and $86$ results give $\mathcal{F}_{\mathrm{obj}} \geq 0.87$, while for SFB method showed in Figure~\ref{Fig_6} (b), only $3$ results out of the $100$ trials give $\mathcal{F}_{\mathrm{obj}} \geq 0.87$. Figure~\ref{Fig_6} (b) and (e) display the shape of optimized control field in the interaction picture, Figure~\ref{Fig_6} (c) and (f) display the fidelity distribution in region  $\delta \in 2\pi \times [-10,10]$ MHz and $\kappa\in [0.5,1.5]$.
	
	\subsection{Sensitivity improvement in AC magnetometry}
	\begin{figure}[] 
		\centering 
		\includegraphics[width = 8cm]{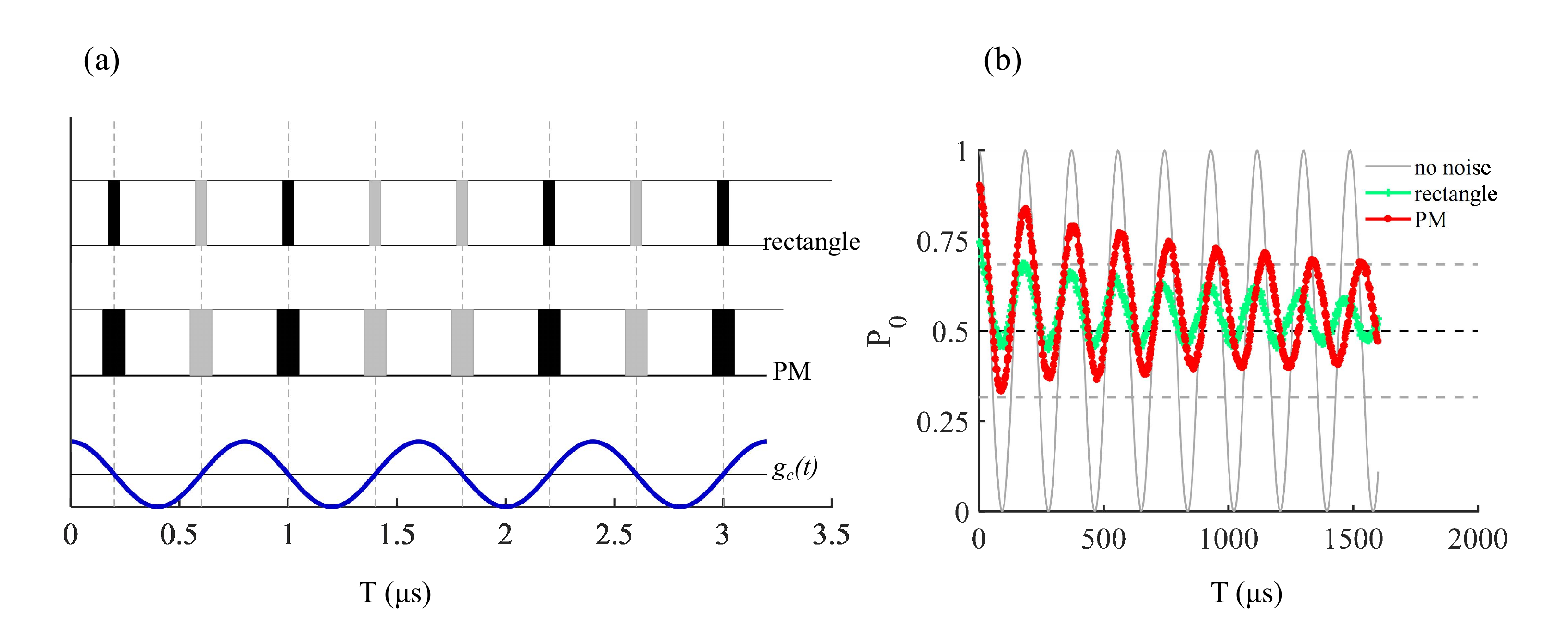}
		\caption{\textbf{(a)} Scheme of the XY-8 pulse and the AC signal to be sensed. The pulse length for rectangular pulse is $T_{\text{pulse}} = 50$ ns and for PM pulse is $T_{\text{pulse}} = 100$ ns. The time separation for rectangular pulse is $\tau = 350$ ns and for PM pulse is  $\tau = 300$ ns The frequency of the AC signal is $\omega_s = \pi/(T_{\text{pulse}}+\tau)= 2.5\pi$ MHz. \textbf{(b)} Simulation results of the population of $\ket{0}$ of the NV center ensemble using different XY-8 pulses. Red line with point marks: population under optimized pulse given by the B-PM method with pulse length $T_{\text{pulse}} = 100$ ns. Green line with cross marks: population under rectangular $\pi$ pulse with pulse length $T_{\text{pulse}} = 50$ ns. Gray line: population under rectangular $\pi$ pulse with pulse length $T_{\text{pulse}} = 100$ ns without considering the inhomogeneous broadening or the dynamic noise term. }
		\label{Fig_7}
	\end{figure}
	The optimal control field of Pauli-X and Pauli-Y gate given by the B-PM method are showed in Fig.(a), with $N_{\mathrm{D}} = 2$, $\boldsymbol{a} = \{0.0583, 0.0046\}$, $\boldsymbol{b} = \{0.0844, 0.1493\}$ and $\boldsymbol{\nu} = \{0.0307, 0.0413 \}$. Figure~\ref{Fig_7} (b) shows the population of the sensor in $\ket{0}$, measured at every terminal of single XY-8 period, such the time interval between two data points is $3.2 \ \mu \mathrm{s}$. $1000$ measurement is made under different $\delta$ obey Gaussian distribution with zero mean value and FWHM = $2\pi\times 26.5 \ \mathrm{MHz}$. Because of inhomogeneous broaden of $\delta$ as well as the dynamic noise $\delta(t)$, the population of sensor decays with time, featured by $T_2\approx 180 \ \mu \mathrm{s}$ for rectangular XY-8 pulse and $T_2\approx 1500 \ \mu \mathrm{s}$ for PM XY-8 pulse respectively. 
	%Accordingly, the sensitivity to the AC field which proportional to $\exp\left(-8(\tau_{\text{pulse}}+T_{\text{pulse}})/T_2\right)$\cite{genovEfficientRobustSignal2020a} is also improved by $2\%$ comparing to the standard rectangular XY-8 sequence.
	%%%%%%%%%%%%%%%%%%%%%%%%%%%%%%%%%%%%%%%%%%
	\section{Discussion}
	Our work shows the Bayesian based estimation model can be effectively applied to estimating the fidelity of the state transformation and the gate construction of NV center under wide inhomogeneous broadening noise. The estimation accuracy as well as the time efficiency make it an ideal tool to constitute a new practical optimization method, which we denote as Bayesian estimation phase-modulated (B-PM) method. B-PM method can be combined with various of search algorithms, i.e., the direct search algorithm\cite{canevaChoppedRandombasisQuantum2011,mullerOneDecadeQuantum2022}, the genetic algorithm\cite{katochReviewGeneticAlgorithm2021} and the gradient based algorithm\cite{khanejaOptimalControlCoupled2005,machnesComparingOptimizingBenchmarking2011,lucarelliQuantumOptimalControl2018,sorensenQuantumOptimalControl2018}. Here we adopt the widely used Nelder-Mead direct search method\cite{nelderSimplexMethodFunction1965a}, related optimization tools is easily available on most of common programming platforms. The Nelder-Mead method seeks the best parameter point by successively construct new points to replace the worst point following a heuristic simplex approach. Consequently the objective function need to be frequently called to rank the points during the searching process, which consume the majority of the total processing time. The hybrid Bayesian estimation phase-modulated (B-PM) method reduce the processing time from two aspects. On the one hand, by setting the predict function as the objective function, the time needed to compute the objective function value once is reduced, as showed in Figure~\ref{Fig_4}. On the other hand, the phase-modulated method can significantly decrease the total number that objective function are called, by introducing a simpler landscape of the parameter space. Taken these two factors together, the advantage of the B-PM method becomes obvious compared to the commonly used standard Fourier base (SFB) method. To be specific, compared to the best result given by the SFB method, the B-PM method raises the optimal value of objective function from $0.894$ to $0.905$, with more than $90\%$ decrease of the average time consumption. Moreover, among $100$ results begin from random initial parameters, only $1$ results of SFB ($N_\text{D} = 2$, $M\times N = 16$) method give $\mathcal{F}_{\mathrm{obj}} \geq 0.89$, while $42$ results of B-PM ($N_\text{D} = 1$, $n = 9$) method give $\mathcal{F}_{\mathrm{obj}} \geq 0.9$, indicating the B-PM method is much more robust to reach a fine result when the amount of total trials is limited. When making a comparison between B-PM and PM method, people may query that the improvement induced by Bayesian estimation model is not obvious. This is true for the current system studied in this article, and needs to be tested in more physical models and systems.
	However, we stress here that based on the results showed by Figure~\ref{Fig_5}, the B-PM method guarantees a higher value of objective function when the computation resources are on the same scale compared with PM method. This is crucial when the computation resource is severely stressed, and when the control performance is sensitive to the precise value of the objective function.
	
	We further display the utility of B-PM method in NV center based AC magnetometry, especially in high frequency cases where the adiabatic strategy is not viable. The numerical simulation result shows the optimized pulse achieves a four-fold extension of the coherence time $T_2$ and a two fold improvement of the sensitivity compared to the conventional rectangular $\pi$ pulses with the same maximum amplitude. Similar optimization strategy can applied to other sensing cases, e.g., the spin bath driving \cite{knowlesObservingBulkDiamond2014,delangeControllingQuantumDynamics2012,bauchUltralongDephasingTimes2018} process for DC magnetometry sensing. 
	
	One important potential application area of B-PM method is the closed-loop optimization\cite{eggerAdaptiveHybridOptimal2014a,accantoRapidRobustControl2017,yangProbeOptimizationQuantum2020} for complex system such as many-body\cite{doriaOptimalControlTechnique2011}and many electron systems\cite{castroControllingDynamicsManyElectron2012}. In these cases numerical simulations fail to describe the system accurately due to 
	a lack of information of the complex system, and the experimental results are directly used as the value of objective function in optimization process. The advantage of B-PM method in reducing the requisite total sample number during the process will be more prominent in such conditions since the time needed to experimentally obtain one sample data is generally much longer compared to the simulation process. Besides NV center, the B-PM method is also expected be applied to optimization of other prevailing quantum platforms such as trapped ions\cite{zhangNOONStatesNine2018,monz14QubitEntanglementCreation2011,singerColloquiumTrappedIons2010a}, cold atoms and superconducting qubits\cite{wattsOptimizingArbitraryPerfect2015,goerzOptimizingArbitraryPerfect2015b}.

\bibliographystyle{aps}
\bibliography{Bayesian0215}

\begin{thebibliography}{10}
\makeatletter
\providecommand \href@noop[0]{\@secondoftwo}%

\bibitem{taylorHighsensitivityDiamondMagnetometer2008} J.~M. Taylor,
  P.~Cappellaro, L.~Childress, L.~Jiang, D.~Budker, P.~R. Hemmer, A.~Yacoby,
  R.~Walsworth, and M.~D. Lukin, \textit{High-Sensitivity Diamond Magnetometer
  with Nanoscale Resolution}, \href{https://doi.org/10.1038/nphys1075}{Nature
  Physics \textbf{4}, 810 (2008)}.

\bibitem{rondinMagnetometryNitrogenvacancyDefects2014} L.~Rondin, J.-P.
  Tetienne, T.~Hingant, J.-F. Roch, P.~Maletinsky, and V.~Jacques,
  \textit{Magnetometry with Nitrogen-Vacancy Defects in Diamond},
  \href{https://doi.org/10.1088/0034-4885/77/5/056503}{Reports on Progress in
  Physics \textbf{77}, 056503 (2014)}.

\bibitem{casolaProbingCondensedMatter2018} F.~Casola, T.~{van der Sar}, and
  A.~Yacoby, \textit{Probing Condensed Matter Physics with Magnetometry Based
  on Nitrogen-Vacancy Centres in Diamond},
  \href{https://doi.org/10.1038/natrevmats.2017.88}{Nature Reviews Materials
  \textbf{3}, 17088 (2018)}.

\bibitem{doldeElectricfieldSensingUsing2011a} F.~Dolde, H.~Fedder, M.~W.
  Doherty, T.~N{\"o}bauer, F.~Rempp, G.~Balasubramanian, T.~Wolf, F.~Reinhard,
  L.~C.~L. Hollenberg, F.~Jelezko, and J.~Wrachtrup, \textit{Electric-Field
  Sensing Using Single Diamond Spins},
  \href{https://doi.org/10.1038/nphys1969}{Nature Physics \textbf{7}, 459
  (2011)}.

\bibitem{neumannHighPrecisionNanoscaleTemperature2013a} P.~Neumann, I.~Jakobi,
  F.~Dolde, C.~Burk, R.~Reuter, G.~Waldherr, J.~Honert, T.~Wolf, A.~Brunner,
  J.~H. Shim, D.~Suter, H.~Sumiya, J.~Isoya, and J.~Wrachtrup,
  \textit{High-{{Precision Nanoscale Temperature Sensing Using Single Defects}}
  in {{Diamond}}}, \href{https://doi.org/10.1021/nl401216y}{Nano Letters
  \textbf{13}, 2738 (2013)}.

\bibitem{PhysRevApplied.10.034009} K.~Hayashi, Y.~Matsuzaki, T.~Taniguchi,
  T.~{Shimo-Oka}, I.~Nakamura, S.~Onoda, T.~Ohshima, H.~Morishita, M.~Fujiwara,
  S.~Saito, and N.~Mizuochi, \textit{Optimization of Temperature Sensitivity
  Using the Optically Detected Magnetic-Resonance Spectrum of a
  Nitrogen-Vacancy Center Ensemble},
  \href{https://doi.org/10.1103/PhysRevApplied.10.034009}{Physical Review
  Applied \textbf{10}, 034009 (2018)}.

\bibitem{dohertyElectronicPropertiesMetrology2014d} M.~W. Doherty, V.~V.
  Struzhkin, D.~A. Simpson, L.~P. McGuinness, Y.~Meng, A.~Stacey, T.~J. Karle,
  R.~J. Hemley, N.~B. Manson, L.~C.~L. Hollenberg, and S.~Prawer,
  \textit{Electronic {{Properties}} and {{Metrology Applications}} of the
  {{Diamond}} \$\{\textbackslash
  mathrm\{\vphantom{\}\}}{{NV}}\vphantom\{\}\vphantom\{\}\^\{\textbackslash
  ensuremath\{-\}\}\$ {{Center}} under {{Pressure}}},
  \href{https://doi.org/10.1103/PhysRevLett.112.047601}{Physical Review Letters
  \textbf{112}, 047601 (2014)}.

\bibitem{wangPicoteslaMagnetometryMicrowave2022} Z.~Wang, F.~Kong, P.~Zhao,
  Z.~Huang, P.~Yu, Y.~Wang, F.~Shi, and J.~Du, \textit{Picotesla Magnetometry
  of Microwave Fields with Diamond Sensors},
  \href{https://doi.org/10.1126/sciadv.abq8158}{Science Advances \textbf{8},
  eabq8158 (2022)}.

\bibitem{schmittSubmillihertzMagneticSpectroscopy2017} S.~Schmitt, T.~Gefen,
  F.~M. St{\"u}rner, T.~Unden, G.~Wolff, C.~M{\"u}ller, J.~Scheuer,
  B.~Naydenov, M.~Markham, S.~Pezzagna, J.~Meijer, I.~Schwarz, M.~Plenio,
  A.~Retzker, L.~P. McGuinness, and F.~Jelezko, \textit{Submillihertz Magnetic
  Spectroscopy Performed with a Nanoscale Quantum Sensor},
  \href{https://doi.org/10.1126/science.aam5532}{Science \textbf{356}, 832
  (2017)}.

\bibitem{millerSpinenhancedNanodiamondBiosensing2020} B.~S. Miller, L.~Bezinge,
  H.~D. Gliddon, D.~Huang, G.~Dold, E.~R. Gray, J.~Heaney, P.~J. Dobson,
  E.~Nastouli, J.~J.~L. Morton, and R.~A. McKendry, \textit{Spin-Enhanced
  Nanodiamond Biosensing for Ultrasensitive Diagnostics},
  \href{https://doi.org/10.1038/s41586-020-2917-1}{Nature \textbf{587}, 588
  (2020)}.

\bibitem{liSARSCoV2QuantumSensor2022} C.~Li, R.~Soleyman, M.~Kohandel, and
  P.~Cappellaro, \textit{{{SARS-CoV-2 Quantum Sensor Based}} on
  {{Nitrogen-Vacancy Centers}} in {{Diamond}}},
  \href{https://doi.org/10.1021/acs.nanolett.1c02868}{Nano Letters \textbf{22},
  43 (2022)}.

\bibitem{araiMillimetrescaleMagnetocardiographyLiving2022} K.~Arai,
  A.~Kuwahata, D.~Nishitani, I.~Fujisaki, R.~Matsuki, Y.~Nishio, Z.~Xin,
  X.~Cao, Y.~Hatano, S.~Onoda, C.~Shinei, M.~Miyakawa, T.~Taniguchi,
  M.~Yamazaki, T.~Teraji, T.~Ohshima, M.~Hatano, M.~Sekino, and T.~Iwasaki,
  \textit{Millimetre-Scale Magnetocardiography of Living Rats with
  Thoracotomy},
  \href{https://doi.org/10.1038/s42005-022-00978-0}{Communications Physics
  \textbf{5}, 200 (2022)}.

\bibitem{chenImmunomagneticMicroscopyTumor2022} S.~Chen, W.~Li, X.~Zheng,
  P.~Yu, P.~Wang, Z.~Sun, Y.~Xu, D.~Jiao, X.~Ye, M.~Cai, M.~Shen, M.~Wang,
  Q.~Zhang, F.~Kong, Y.~Wang, J.~He, H.~Wei, F.~Shi, and J.~Du,
  \textit{Immunomagnetic Microscopy of Tumor Tissues Using Sensors in Diamond},
  \href{https://doi.org/10.1073/pnas.2118876119}{Proceedings of the National
  Academy of Sciences of the United States of America \textbf{119}, e2118876119
  (2022)}.

\bibitem{wangRandomizationPulsePhases2019} Z.-Y. Wang, J.~E. Lang, S.~Schmitt,
  J.~Lang, J.~Casanova, L.~McGuinness, T.~S. Monteiro, F.~Jelezko, and M.~B.
  Plenio, \textit{Randomization of {{Pulse Phases}} for {{Unambiguous}} and
  {{Robust Quantum Sensing}}},
  \href{https://doi.org/10.1103/PhysRevLett.122.200403}{Physical Review Letters
  \textbf{122}, 200403 (2019)}.

\bibitem{macquarrieContinuousDynamicalDecoupling2015} E.~R. MacQuarrie, T.~A.
  Gosavi, S.~A. Bhave, and G.~D. Fuchs, \textit{Continuous Dynamical Decoupling
  of a Single Diamond Nitrogen-Vacancy Center Spin with a Mechanical
  Resonator}, \href{https://doi.org/10.1103/PhysRevB.92.224419}{Physical Review
  B \textbf{92}, 224419 (2015)}.

\bibitem{caoProtectingQuantumSpin2020b} Q.-Y. Cao, P.-C. Yang, M.-S. Gong,
  M.~Yu, A.~Retzker, M.~Plenio, C.~M{\"u}ller, N.~Tomek, B.~Naydenov,
  L.~McGuinness, F.~Jelezko, and J.-M. Cai, \textit{Protecting {{Quantum Spin
  Coherence}} of {{Nanodiamonds}} in {{Living Cells}}},
  \href{https://doi.org/10.1103/PhysRevApplied.13.024021}{Physical Review
  Applied \textbf{13}, 024021 (2020)}.

\bibitem{farfurnikOptimizingDynamicalDecoupling2015a} D.~Farfurnik, A.~Jarmola,
  L.~M. Pham, Z.~H. Wang, V.~V. Dobrovitski, R.~L. Walsworth, D.~Budker, and
  N.~{Bar-Gill}, \textit{Optimizing a Dynamical Decoupling Protocol for
  Solid-State Electronic Spin Ensembles in Diamond},
  \href{https://doi.org/10.1103/PhysRevB.92.060301}{Physical Review B
  \textbf{92}, 060301 (2015)}.

\bibitem{genovEfficientRobustSignal2020a} G.~T. Genov, Y.~{Ben-Shalom},
  F.~Jelezko, A.~Retzker, and N.~{Bar-Gill}, \textit{Efficient and Robust
  Signal Sensing by Sequences of Adiabatic Chirped Pulses},
  \href{https://doi.org/10.1103/PhysRevResearch.2.033216}{Physical Review
  Research \textbf{2}, 033216 (2020)}.

\bibitem{poulsenOptimalControlNitrogenvacancy2022} A.~F.~L. Poulsen, J.~D.
  Clement, J.~L. Webb, R.~H. Jensen, L.~Troise, K.~{Berg-S{\o}rensen}, A.~Huck,
  and U.~L. Andersen, \textit{Optimal Control of a Nitrogen-Vacancy Spin
  Ensemble in Diamond for Sensing in the Pulsed Domain},
  \href{https://doi.org/10.1103/PhysRevB.106.014202}{Physical Review B
  \textbf{106}, 014202 (2022)}.

\bibitem{brochuTutorialBayesianOptimization2010} E.~Brochu, V.~M. Cora, and
  N.~{de Freitas}, \textit{A {{Tutorial}} on {{Bayesian Optimization}} of
  {{Expensive Cost Functions}}, with {{Application}} to {{Active User
  Modeling}} and {{Hierarchical Reinforcement Learning}}},
  \href@noop{}{arXiv:1012.2599 [cs]  (2010)}.

\bibitem{shahriariTakingHumanOut2016} B.~Shahriari, K.~Swersky, Z.~Wang, R.~P.
  Adams, and N.~{de Freitas}, \textit{Taking the {{Human Out}} of the {{Loop}}:
  {{A Review}} of {{Bayesian Optimization}}},
  \href{https://doi.org/10.1109/JPROC.2015.2494218}{Proceedings of the IEEE
  \textbf{104}, 148 (2016)}.

\bibitem{zhanExpectedImprovementExpensive2020} D.~Zhan and H.~Xing,
  \textit{Expected Improvement for Expensive Optimization: A Review},
  \href{https://doi.org/10.1007/s10898-020-00923-x}{Journal of Global
  Optimization \textbf{78}, 507 (2020)}.

\bibitem{bauchUltralongDephasingTimes2018} E.~Bauch, C.~A. Hart, J.~M. Schloss,
  M.~J. Turner, J.~F. Barry, P.~Kehayias, S.~Singh, and R.~L. Walsworth,
  \textit{Ultralong {{Dephasing Times}} in {{Solid-State Spin Ensembles}} via
  {{Quantum Control}}},
  \href{https://doi.org/10.1103/PhysRevX.8.031025}{Physical Review X
  \textbf{8}, 031025 (2018)}.

\bibitem{glaserTrainingSchrodingerCat2015} S.~J. Glaser, U.~Boscain,
  T.~Calarco, C.~P. Koch, W.~K{\"o}ckenberger, R.~Kosloff, I.~Kuprov, B.~Luy,
  S.~Schirmer, T.~{Schulte-Herbr{\"u}ggen}, D.~Sugny, and F.~K. Wilhelm,
  \textit{Training {{Schr\"odinger}}'s Cat: Quantum Optimal Control},
  \href{https://doi.org/10.1140/epjd/e2015-60464-1}{The European Physical
  Journal D \textbf{69}, 279 (2015)}.

\bibitem{kochQuantumOptimalControl2022} C.~P. Koch, U.~Boscain, T.~Calarco,
  G.~Dirr, S.~Filipp, S.~J. Glaser, R.~Kosloff, S.~Montangero,
  T.~{Schulte-Herbr{\"u}ggen}, D.~Sugny, and F.~K. Wilhelm, \textit{Quantum
  Optimal Control in Quantum Technologies. {{Strategic}} Report on Current
  Status, Visions and Goals for Research in {{Europe}}},
  \href{https://doi.org/10.1140/epjqt/s40507-022-00138-x}{EPJ Quantum
  Technology \textbf{9}, 19 (2022)}.

\bibitem{accantoRapidRobustControl2017} N.~Accanto, P.~M. {de Roque},
  M.~{Galvan-Sosa}, S.~Christodoulou, I.~Moreels, and N.~F. {van Hulst},
  \textit{Rapid and Robust Control of Single Quantum Dots},
  \href{https://doi.org/10.1038/lsa.2016.239}{Light: Science \& Applications
  \textbf{6}, e16239 (2017)}.

\bibitem{yangProbeOptimizationQuantum2020} X.~Yang, J.~Thompson, Z.~Wu, M.~Gu,
  X.~Peng, and J.~Du, \textit{Probe Optimization for Quantum Metrology via
  Closed-Loop Learning Control},
  \href{https://doi.org/10.1038/s41534-020-00292-z}{npj Quantum Information
  \textbf{6}, 1 (2020)}.

\bibitem{eggerAdaptiveHybridOptimal2014a} D.~J. Egger and F.~K. Wilhelm,
  \textit{Adaptive {{Hybrid Optimal Quantum Control}} for {{Imprecisely
  Characterized Systems}}},
  \href{https://doi.org/10.1103/PhysRevLett.112.240503}{Physical Review Letters
  \textbf{112}, 240503 (2014)}.

\bibitem{jelezkoSingleDefectCentres2006} F.~Jelezko and J.~Wrachtrup,
  \textit{Single Defect Centres in Diamond: {{A}} Review}, \href@noop{}{physica
  status solidi (a) \textbf{203}, 3207 (2006)}.

\bibitem{tianQuantumOptimalControl2020} J.~Tian, H.~Liu, Y.~Liu, P.~Yang,
  R.~Betzholz, R.~S. Said, F.~Jelezko, and J.~Cai, \textit{Quantum Optimal
  Control Using Phase-Modulated Driving Fields},
  \href{https://doi.org/10.1103/PhysRevA.102.043707}{Physical Review A
  \textbf{102}, 043707 (2020)}.

\bibitem{simpsonMetamodelsComputerbasedEngineering2001} T.~Simpson,
  J.~Poplinski, P.~N. Koch, and J.~Allen, \textit{Metamodels for
  {{Computer-based Engineering Design}}: {{Survey}} and Recommendations},
  \href{https://doi.org/10.1007/PL00007198}{Engineering with Computers
  \textbf{17}, 129 (2001)}.

\bibitem{wangReviewMetamodelingTechniques2006} G.~G. Wang and S.~Shan,
  \textit{Review of {{Metamodeling Techniques}} in {{Support}} of {{Engineering
  Design Optimization}}}, \href{https://doi.org/10.1115/1.2429697}{Journal of
  Mechanical Design \textbf{129}, 370 (2006)}.

\bibitem{r.r.bartonMetamodelingStateArt11} {R. R. Barton}, in
  \textit{Proceedings of {{Winter Simulation Conference}}},  (, 11).

\bibitem{vandeschootBayesianStatisticsModelling2021} R.~{van de Schoot},
  S.~Depaoli, R.~King, B.~Kramer, K.~M{\"a}rtens, M.~G. Tadesse, M.~Vannucci,
  A.~Gelman, D.~Veen, J.~Willemsen, and C.~Yau, \textit{Bayesian Statistics and
  Modelling}, \href{https://doi.org/10.1038/s43586-020-00001-2}{Nature Reviews
  Methods Primers \textbf{1}, 1 (2021)}.

\bibitem{jeromesacksDesignAnalysisComputer1989} {Jerome Sacks}, {William J.
  Welch}, {Toby J. Mitchell}, and {Henry P. Wynn}, \textit{Design and
  {{Analysis}} of {{Computer Experiments}}},
  \href{https://doi.org/10.1214/ss/1177012413}{Statistical Science \textbf{4},
  409 (1989)}.

\bibitem{jonesEfficientGlobalOptimization1998} D.~R. Jones and M.~Schonlau,
  \textit{Efficient {{Global Optimization}} of {{Expensive Black-Box
  Functions}}}, \href@noop{}{Journal of Global optimization 38 (1998)}.

\bibitem{matheronPrinciplesGeostatistics1963} G.~Matheron, \textit{Principles
  of Geostatistics},
  \href{https://doi.org/10.2113/gsecongeo.58.8.1246}{Economic Geology
  \textbf{58}, 1246 (1963)}.

\bibitem{ngBayesianKrigingAnalysis2012} S.~H. Ng and J.~Yin, \textit{Bayesian
  {{Kriging Analysis}} and {{Design}} for {{Stochastic Simulations}}},
  \href{https://doi.org/10.1145/2331140.2331145}{Acm Transactions on Modeling
  and Computer Simulation \textbf{22}, 17 (2012)}.

\bibitem{steinInterpolationSpatialData1999} M.~L. Stein,
  \href@noop{}{\textit{Interpolation of Spatial Data: Some Theory for
  Kriging}},  ({Springer Science \& Business Media}, 1999).

\bibitem{martinUseKrigingModels2005a} J.~D. Martin and T.~W. Simpson,
  \textit{Use of {{Kriging Models}} to {{Approximate Deterministic Computer
  Models}}}, \href{https://doi.org/10.2514/1.8650}{AIAA Journal \textbf{43},
  853 (2005)}.

\bibitem{currinBayesianApproachDesign1988} C.~Currin, T.~Mitchell, M.~Morris,
  and D.~Ylvisaker, \textit{A {{Bayesian}} Approach to the Design and Analysis
  of Computer Experiments}, Technical Report ORNL-6498, {Oak Ridge National
  Lab., TN (USA)} (1988).

\bibitem{currinBayesianPredictionDeterministic1991} C.~Currin, T.~Mitchell,
  M.~Morris, and D.~Ylvisaker, \textit{Bayesian {{Prediction}} of
  {{Deterministic Functions}}, with {{Applications}} to the {{Design}} and
  {{Analysis}} of {{Computer Experiments}}},
  \href{https://doi.org/10.2307/2290511}{Journal of the American Statistical
  Association \textbf{86}, 953 (1991)}.

\bibitem{morrisBayesianDesignAnalysis1993} M.~D. Morris, T.~J. Mitchell, and
  D.~Ylvisaker, \textit{Bayesian {{Design}} and {{Analysis}} of {{Computer
  Experiments}}: {{Use}} of {{Derivatives}} in {{Surface Prediction}}},
  \href{https://doi.org/10.1080/00401706.1993.10485320}{Technometrics
  \textbf{35}, 243 (1993)}.

\bibitem{kleijnenKrigingMetamodelingSimulation2009} J.~P. Kleijnen,
  \textit{Kriging Metamodeling in Simulation: {{A}} Review},
  \href{https://doi.org/10.1016/j.ejor.2007.10.013}{European Journal of
  Operational Research \textbf{192}, 707 (2009)}.

\bibitem{canevaChoppedRandombasisQuantum2011} T.~Caneva, T.~Calarco, and
  S.~Montangero, \textit{Chopped Random-Basis Quantum Optimization},
  \href{https://doi.org/10.1103/PhysRevA.84.022326}{Physical Review A
  \textbf{84}, 022326 (2011)}.

\bibitem{mullerOneDecadeQuantum2022} M.~M. M{\"u}ller, R.~S. Said, F.~Jelezko,
  T.~Calarco, and S.~Montangero, \textit{One Decade of Quantum Optimal Control
  in the Chopped Random Basis},
  \href{https://doi.org/10.1088/1361-6633/ac723c}{Reports on Progress in
  Physics \textbf{85}, 076001 (2022)}.

\bibitem{katochReviewGeneticAlgorithm2021} S.~Katoch, S.~S. Chauhan, and
  V.~Kumar, \textit{A Review on Genetic Algorithm: Past, Present, and Future},
  \href{https://doi.org/10.1007/s11042-020-10139-6}{Multimedia Tools and
  Applications \textbf{80}, 8091 (2021)}.

\bibitem{khanejaOptimalControlCoupled2005} N.~Khaneja, T.~Reiss, C.~Kehlet,
  T.~{Schulte-Herbr{\"u}ggen}, and S.~J. Glaser, \textit{Optimal Control of
  Coupled Spin Dynamics: Design of {{NMR}} Pulse Sequences by Gradient Ascent
  Algorithms}, \href{https://doi.org/10.1016/j.jmr.2004.11.004}{Journal of
  Magnetic Resonance \textbf{172}, 296 (2005)}.

\bibitem{machnesComparingOptimizingBenchmarking2011} S.~Machnes, U.~Sander,
  S.~J. Glaser, P.~{de Fouqui{\`e}res}, A.~Gruslys, S.~Schirmer, and
  T.~{Schulte-Herbr{\"u}ggen}, \textit{Comparing, Optimizing, and Benchmarking
  Quantum-Control Algorithms in a Unifying Programming Framework},
  \href{https://doi.org/10.1103/PhysRevA.84.022305}{Physical Review A
  \textbf{84}, 022305 (2011)}.

\bibitem{lucarelliQuantumOptimalControl2018} D.~Lucarelli, \textit{Quantum
  Optimal Control via Gradient Ascent in Function Space and the Time-Bandwidth
  Quantum Speed Limit},
  \href{https://doi.org/10.1103/PhysRevA.97.062346}{Physical Review A
  \textbf{97}, 062346 (2018)}.

\bibitem{sorensenQuantumOptimalControl2018} J.~J. W.~H. S{\o}rensen, M.~O.
  Aranburu, T.~Heinzel, and J.~F. Sherson, \textit{Quantum Optimal Control in a
  Chopped Basis: {{Applications}} in Control of {{Bose-Einstein}} Condensates},
  \href{https://doi.org/10.1103/PhysRevA.98.022119}{Physical Review A
  \textbf{98}, 022119 (2018)}.

\bibitem{nelderSimplexMethodFunction1965a} J.~A. Nelder and R.~Mead, \textit{A
  {{Simplex Method}} for {{Function Minimization}}},
  \href{https://doi.org/10.1093/comjnl/7.4.308}{The Computer Journal
  \textbf{7}, 308 (1965)}.

\bibitem{knowlesObservingBulkDiamond2014} H.~S. Knowles, D.~M. Kara, and
  M.~Atat{\"u}re, \textit{Observing Bulk Diamond Spin Coherence in High-Purity
  Nanodiamonds}, \href{https://doi.org/10.1038/nmat3805}{Nature Materials
  \textbf{13}, 21 (2014)}.

\bibitem{delangeControllingQuantumDynamics2012} G.~{de Lange}, T.~{van der
  Sar}, M.~Blok, Z.-H. Wang, V.~Dobrovitski, and R.~Hanson, \textit{Controlling
  the Quantum Dynamics of a Mesoscopic Spin Bath in Diamond},
  \href{https://doi.org/10.1038/srep00382}{Scientific Reports \textbf{2}, 382
  (2012)}.

\bibitem{doriaOptimalControlTechnique2011} P.~Doria, T.~Calarco, and
  S.~Montangero, \textit{Optimal Control Technique for Many-Body Quantum
  Dynamics}, \href{https://doi.org/10.1103/PhysRevLett.106.190501}{Physical
  Review Letters \textbf{106}, 190501 (2011)}.

\bibitem{castroControllingDynamicsManyElectron2012} A.~Castro, J.~Werschnik,
  and E.~K.~U. Gross, \textit{Controlling the {{Dynamics}} of {{Many-Electron
  Systems}} from {{First Principles}}: {{A Combination}} of {{Optimal Control}}
  and {{Time-Dependent Density-Functional Theory}}},
  \href{https://doi.org/10.1103/PhysRevLett.109.153603}{Physical Review Letters
  \textbf{109}, 153603 (2012)}.

\bibitem{zhangNOONStatesNine2018} J.~Zhang, M.~Um, D.~Lv, J.-N. Zhang, L.-M.
  Duan, and K.~Kim, \textit{{{NOON States}} of {{Nine Quantized Vibrations}} in
  {{Two Radial Modes}} of a {{Trapped Ion}}},
  \href{https://doi.org/10.1103/PhysRevLett.121.160502}{Physical Review Letters
  \textbf{121}, 160502 (2018)}.

\bibitem{monz14QubitEntanglementCreation2011} T.~Monz, P.~Schindler, J.~T.
  Barreiro, M.~Chwalla, D.~Nigg, W.~A. Coish, M.~Harlander, W.~H{\"a}nsel,
  M.~Hennrich, and R.~Blatt, \textit{14-{{Qubit Entanglement}}: {{Creation}}
  and {{Coherence}}},
  \href{https://doi.org/10.1103/PhysRevLett.106.130506}{Physical Review Letters
  \textbf{106}, 130506 (2011)}.

\bibitem{singerColloquiumTrappedIons2010a} K.~Singer, U.~Poschinger, M.~Murphy,
  P.~Ivanov, F.~Ziesel, T.~Calarco, and F.~{Schmidt-Kaler}, \textit{Colloquium:
  {{Trapped}} Ions as Quantum Bits: {{Essential}} Numerical Tools},
  \href{https://doi.org/10.1103/RevModPhys.82.2609}{Reviews of Modern Physics
  \textbf{82}, 2609 (2010)}.

\bibitem{wattsOptimizingArbitraryPerfect2015} P.~Watts, J.~Vala, M.~M.
  M{\"u}ller, T.~Calarco, K.~B. Whaley, D.~M. Reich, M.~H. Goerz, and C.~P.
  Koch, \textit{Optimizing for an Arbitrary Perfect Entangler. {{I}}.
  {{Functionals}}}, \href{https://doi.org/10.1103/PhysRevA.91.062306}{Physical
  Review A \textbf{91}, 062306 (2015)}.

\bibitem{goerzOptimizingArbitraryPerfect2015b} M.~H. Goerz, G.~Gualdi, D.~M.
  Reich, C.~P. Koch, F.~Motzoi, K.~B. Whaley, J.~Vala, M.~M. M{\"u}ller,
  S.~Montangero, and T.~Calarco, \textit{Optimizing for an Arbitrary Perfect
  Entangler. {{II}}. {{Application}}},
  \href{https://doi.org/10.1103/PhysRevA.91.062307}{Physical Review A
  \textbf{91}, 062307 (2015)}.

\end{thebibliography}
	
\end{document}